\documentclass[a4paper,12pt]{spieman}  
\usepackage{amsmath,amsfonts,amssymb}
\usepackage{graphicx}
\usepackage{setspace}
\usepackage{tocloft}
\usepackage{booktabs}
\usepackage{lineno}

\title{Recent developments in Laue lens manufacturing and their impact on imaging performance}

\author[a,b]{Lisa~Ferro}
\author[b,c]{Enrico~Virgilli}
\author[b]{Natalia~Auricchio}
\author[d,e]{Claudio~Ferrari}
\author[b]{Ezio~Caroli}
\author[a,d]{Riccardo~Lolli}
\author[a]{Miguel~F.~Moita}
\author[a,b,f]{Piero~Rosati}
\author[a,b]{Filippo~Frontera}
\author[g]{Mauro~Pucci}
\author[b]{John~B.~Stephen}
\author[a,b,f]{Cristiano~Guidorzi}

\affil[a]{University of Ferrara, Department of Physics and Earth Science, Via G. Saragat 1/C, 44122 Ferrara (FE), Italy}

\affil[b]{INAF/OAS of Bologna, Via Piero Gobetti 93/3, 40129 Bologna (BO), Italy }

\affil[c]{INFN Section of Bologna, Viale Berti Pichat, 6/2, 40127 Bologna (BO), Italy}

\affil[d]{Institute of Materials for Electronics and Magnetism (CNR-IMEM), Parco Area delle Scienze 37/A, 43124 Parma (PR), Italy}

\affil[e]{INAF Osservatorio Astronomico di Brera - Merate, Via E. Bianchi, 46, 23807 Merate (LC), Italy}

\affil[f]{INFN Section of Ferrara, Via G. Saragat 1/C, 44122 Ferrara (FE), Italy}

\affil[g]{National Institute of Optics (CNR-INO),
Largo Enrico Fermi 6, 50125 Florence (FI), Italy }

\cftpagenumbersoff{figure}
\cftpagenumbersoff{table} 
\begin{document} 
\maketitle

\begin{abstract}
\newline
We report on recent progress in the development of Laue lenses for applications in hard X/soft gamma-ray astronomy. Here we focus on the realization of a sector of such a lens made of 11 bent Germanium crystals and describe the technological challenges involved in their positioning and alignment with adhesive-based bonding techniques. The accurate alignment and the uniformity of the curvature of the crystals are critical for achieving optimal X-ray focusing capabilities. We have assessed how the errors of misalignment with respect to the main orientation angles of the crystals affect the point spread function (PSF) of the image diffracted by a single sector. We have corroborated these results with simulations carried out with our physical model of the lens, based on a Monte Carlo ray-tracing technique, adopting the geometrical configuration of the Laue sector, the observed assembly accuracy and the measured curvatures of the crystals. 
An extrapolation of the performances achieved on a single sector to an entire Laue lens based on this model has shown that a PSF with half-power-diameter of 4.8 arcmin can be achieved with current technology. This has the potential to lead to a significant improvement in sensitivity of spectroscopic and polarimetric observations in the 50-600 keV band. 

\end{abstract}


{\noindent \footnotesize\textbf{*}Contact Author: Lisa Ferro,  \linkable{frrlsi@unife.it} }

\begin{spacing}{1.}   
\section{Introduction}\label{intro}
\begin{figure}[t!]
    \centering
    \includegraphics[scale = 1.15]{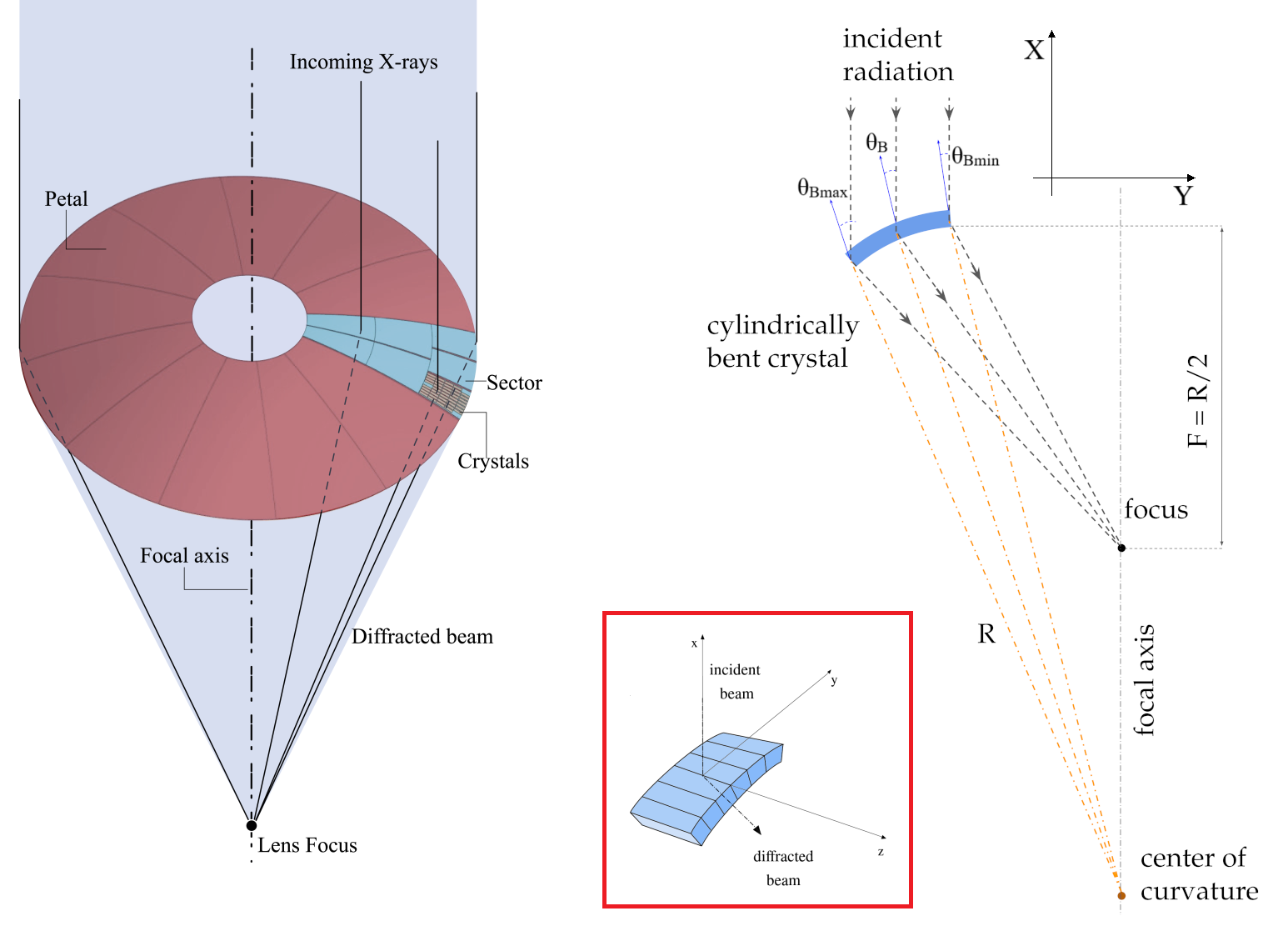}
    \caption{Left: Concept of Laue lens. A number of curved crystal tiles must be arranged in concentric rings and oriented according to the Bragg's law, along an overall curvature with 40 m radius, so as to focus the radiation at the common focal point where a 3-d position-sensitive detector is placed. Right: Radial section of one of the bent crystal (blue arc) which compose the lens and acts as a radiation concentrator.  The red inset shows a 3-d drawing of the crystal including the diffraction planes (black lines) and their orientation.}
    \label{fig:laueconcept}
\end{figure}

\begin{figure}[!t]
\centering
\includegraphics[scale = 1.1]{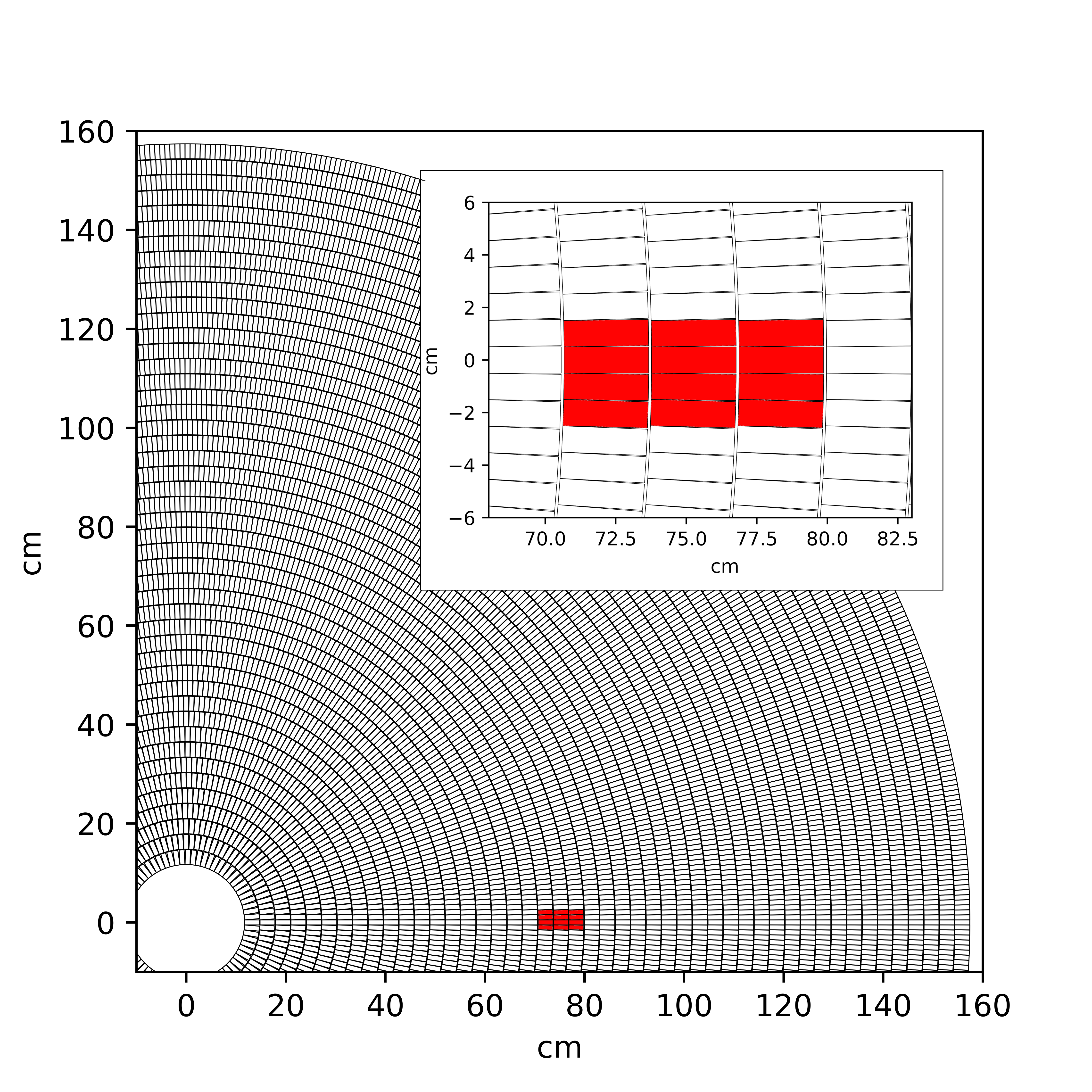}
\caption{Background image: scheme showing $\sim$ 1/4 of the  Laue lens designed for the NFT on board the ASTENA concept mission (internal radius $R_{\rm in}$ = 12~cm, external radius $R_{\rm out}$ = 158~cm). The insert shows a small sector made of 12 crystals, whose prototype model was built and is presented in this work.}
\label{fig:schemeoflens}
\end{figure}

Laue lenses have emerged as a promising and unique technique for focusing hard X-/soft Gamma-ray radiation, offering the ability to achieve high angular resolution while providing large effective areas\cite{frontera2011, Virgilli2016}~. The fundamental principle at the basis of Laue lenses is the single diffraction of high energy photons by a number of crystals arranged in concentric rings (Fig.~\ref{fig:laueconcept}, left)~. By exploiting the diffractive properties of crystals, soft gamma rays can be focused onto a desired focal point, resulting in unparalleled focusing capabilities that can be extended, in principle, up to several hundreds of keV. In particular, our approach is based on the use of curved crystals, which are bent to an external cylindrical curvature in such a way that the average direction of the diffraction planes follow the curvature of the surface. 
In bent crystals, the Bragg's angle $\mathrm{\theta_B}$ between the incoming radiation and the diffraction planes varies continuously, between a maximum ($\mathrm{\theta_{B_{max}}}$) and a minimum ($\mathrm{\theta_{B_{min}}}$), along the surface of the crystal (Fig.~\ref{fig:laueconcept}, right)~. Thus, the diffracted radiation is focused onto an area whose footprint is smaller than the size of the crystal itself.
The practical implementation of Laue lenses has, however, been limited to date by the ability to accurately align a large number of crystals so that their diffracted radiation is concentrated onto a common focal point.
The accurate positioning of thousands of crystals within the lens assembly determines the sharpness of the point spread function (PSF),  which is a key parameter, along with the overall throughput,  when evaluating the scientific performance of the lens in high-energy astrophysics. 
Any deviation from the nominal positioning results in reduced focusing ability, affecting the performance of the instrument.
To address the difficulties encountered in crystal alignment, the choice of the materials that serve as substrates for the crystals plays a crucial role. These materials should possess low thermal expansion, exceptional mechanical stability, and high stiffness. In addition, they should exhibit superior match with the crystal material, in order to ensure an effective bonding between crystal and substrate. Possibly, these materials must be already space-proven or space-qualified. The introduction of these materials has the potential to significantly reduce alignment uncertainties and improve the overall performance of Laue lenses.
From our experience, in order to realize a Laue lens made with numerous optical elements, a modular approach must be adopted. It is therefore necessary to build and align several modules with each other, each consisting of a few dozens of crystals. Within each module, the crystals must be aligned toward the common focus as accurately as possible, since they cannot be further reoriented.
The final objective of our R\&D is to develop 
a Laue lens working in the 50 -- 700~keV energy range as that of the narrow field telescope proposed for the ASTENA mission concept \cite{Frontera2021,Guidorzi2021}~. Such a Laue lens has an aperture diameter of $\sim 3.2$~m and an overall geometric area of 7~m$^2$ (Fig. \ref{fig:schemeoflens}).

In this paper, we report on the latest development in terms of crystals preparation, substrate choice and, in particular, on the realization of a test module of Laue lens made of 12 crystals, aligned with respect to both radial and azimuthal angles.
The achieved positioning accuracy in setting these crystals is then used to predict the PSF of the entire Laue lens using a ray-tracing based physical model of the lens.

\section{Experimental activity}\label{setup}

Goal of this research is to achieve an assembly accuracy which can meet the requirements of a Laue lens for astrophysical applications. In this section, we will describe the relevant components of this endeavour, namely the selected crystals for diffraction, the choice of the substrate used to host the crystals, and the methodological steps of the assembly process in the X-ray facility.

\subsection{Crystals and substrate}
\label{crystals}

\begin{figure}[t!]
    \centering
    \includegraphics[scale = 0.2]{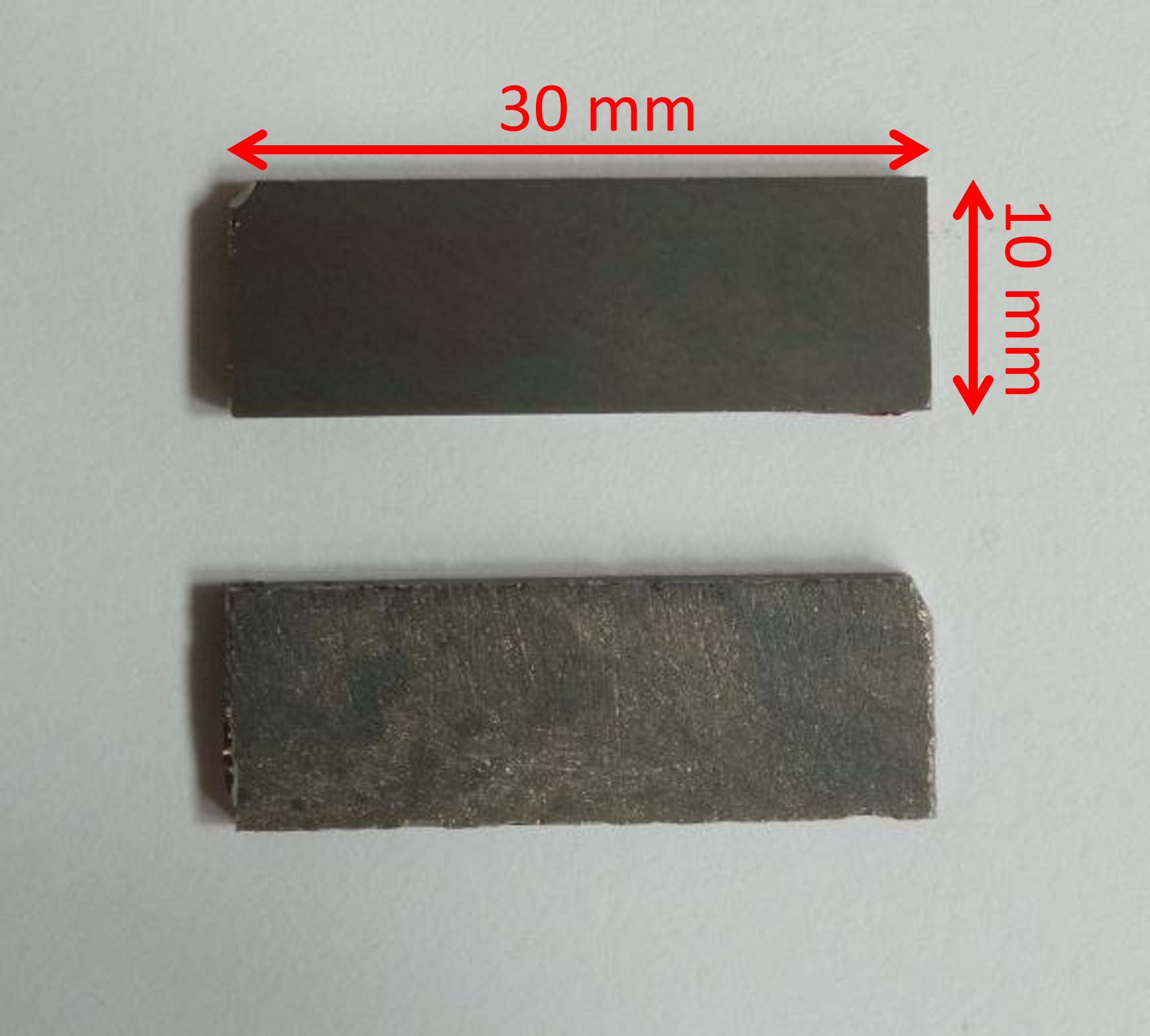}
    \includegraphics[scale = 0.2]{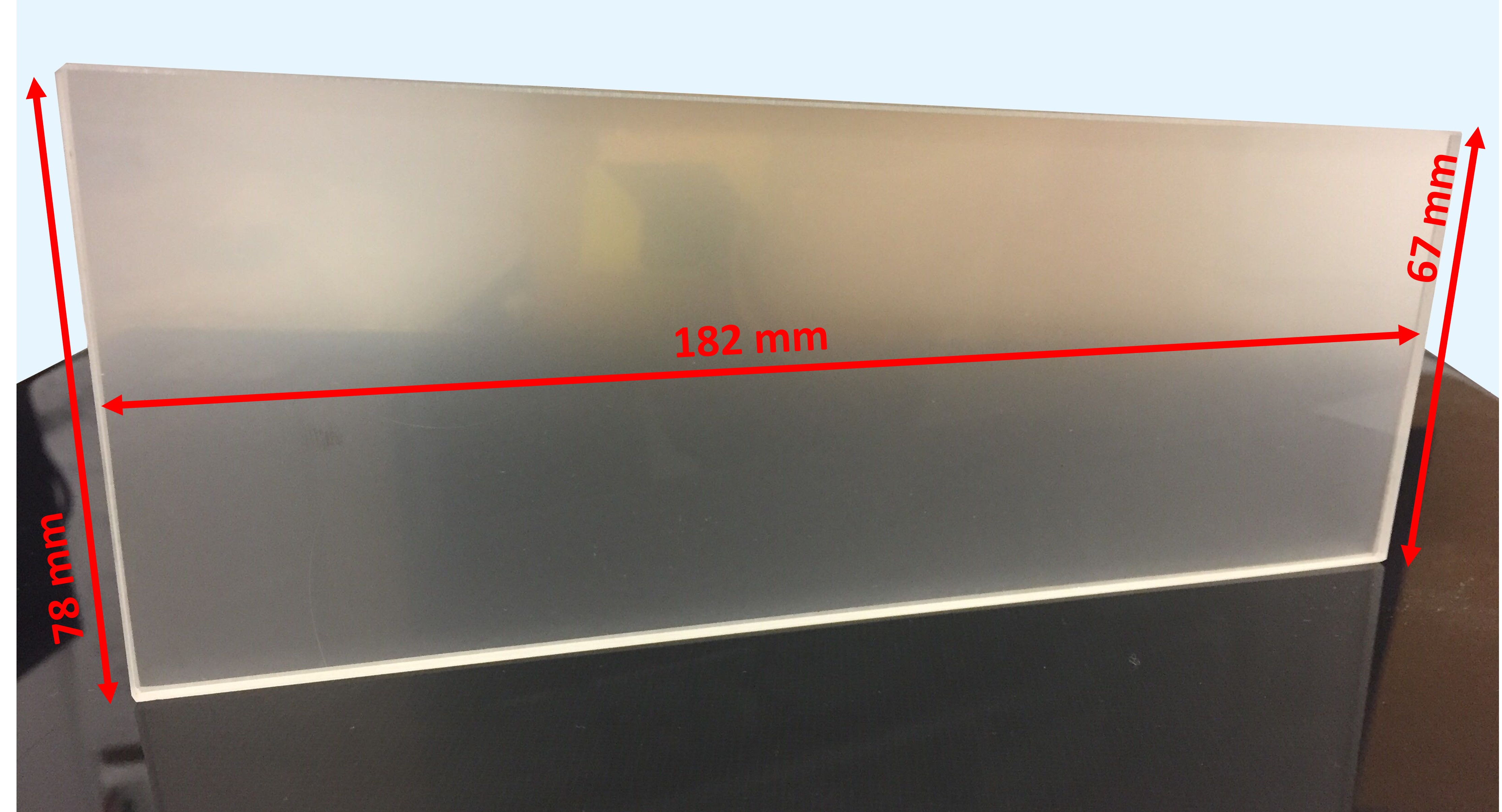}
    \caption{Left: Two Germanium crystal tiles used to build the lens prototype; the bottom crystal shows the polished side. The crystals are bent along the long side. Right: The quartz glass used as substrate for the bonded crystals.}
    \label{fig:substrate_and_crystal}
\end{figure}

The crystals used to realize the sector are made of Germanium tiles (thickness S = 1676 $\div$ 1775 $\mu$m) with cross-section 30~$\times$~10~mm$^2$
(Fig.~\ref{fig:substrate_and_crystal}, left) and diffraction planes (220). The crystal tiles are cut such that the diffraction planes are parallel to the 10~$\times$ S~mm$^2$ faces within a miscut angle $<0.2^{\circ}$. We did not explore in this work the impact of selecting different diffraction planes on the lens performance. This aspect 
will be investigated in future research.

Crystals are bent in a cylindrical shape with a radius of curvature of 40~m. Such a curvature has been obtained through the so-called surface lapping-technique\cite{Ferrari2013,Buffagni2015}~, which consists of a controlled mechanical damaging on one surface of the sample. The procedure provides a highly compressive strain responsible for the convexity appearing on the worked side.
Due to the variability of the amount of material removed by the surface lapping process, the thickness of the crystals is not uniform, as reported in  Table~\ref{tab:crystal_prop}, where also the achieved curvature radius for each crystal tile is reported.
{The planes (220) do not acquire a secondary curvature, as can be expected with other families of planes~\cite{Authier98}~, however the external bending of the crystals allows the X-ray radiation coming from a parallel beam to be} concentrated onto a focal point at a distance equal to half their radius of curvature~\cite{Virgilli2014}~. 

The adhesive used to realize the lens sector is the OP~61~LS by DYMAX, an UV-curable adhesive with a linear shrinkage factor of $0.03$\%. A low value of the adhesive shrinkage is fundamental to reduce unwanted positioning displacements after curing. The substrate we used is made of fused quartz with trapezoid shape (Fig.~\ref{fig:substrate_and_crystal}, right). The choice of quartz as material for the substrate is due to its low coefficient of thermal expansion and for its transparency to the UV light which is required for the adhesive curing process.
One of the two surfaces of the quartz glass has been worked to be flat, while the surface on which the crystals are bonded has been polished to achieve a curvature radius of 40~m to better match the shape of the crystals. This reduces the inhomogeneities in the glue deposition between crystals and substrate.
The substrate is 5~mm thick and the diffraction process is carried out at about 130~keV. At this energy, and based on the density and attenuation coefficient of quartz glass, the beam transmitted through the substrate is about 84\% of the diffracted beam.

\begin{table}[!ht]
\centering
\small
\caption{Curvature radius and thickness of the crystals used to build the prototype. The variation in the reported errors on the curvature radius is due to the different deviations with respect to a perfect curvature for the crystals. Measurements done at CNR-IMEM (Parma) with a Cu-$\mathrm{\alpha}$ X-ray diffractometer. 
}
\begin{tabular}[t]{lcc}
\\
\toprule
Crystal ID&  ~~~~~~~~Curvature Radius~~~~~~~~~   & Thickness\\
  & (m) & ($\mu$m) \\
 \midrule
25A & $\mathrm{40.0 \pm 0.4}$ & $\mathrm{1700 \pm 5}$   \\
26A & $\mathrm{39 \pm 1}$ & $\mathrm{1717 \pm 5}$       \\
26B & $\mathrm{37.8 \pm 0.3}$ & $\mathrm{1775 \pm 5}$   \\
26C & $\mathrm{38.8 \pm 0.8}$ & $\mathrm{1744 \pm 5}$   \\
28B & $\mathrm{40.62 \pm 0.04}$ & $\mathrm{1775 \pm 5}$ \\
29A & $\mathrm{39.08 \pm 0.01}$ & $\mathrm{1732 \pm 5}$ \\
29B & $\mathrm{38.4 \pm 0.9}$ & $\mathrm{1749 \pm 5}$   \\
30A & $\mathrm{38.1 \pm 0.6}$ & $\mathrm{1676 \pm 5}$   \\
31A & $\mathrm{40.6 \pm 0.1}$ & $\mathrm{1745 \pm 5}$   \\
32C & $\mathrm{40.7 \pm 0.7}$ & $\mathrm{1731 \pm 5}$   \\
33B & $\mathrm{40.8 \pm 0.3}$ & $\mathrm{1678 \pm 5}$   \\
\bottomrule
\label{tab:crystal_prop}
\end{tabular}
\end{table}

\subsection{Facility set-up}

\begin{figure}[t!]
    \centering
    \includegraphics[scale = 0.485]{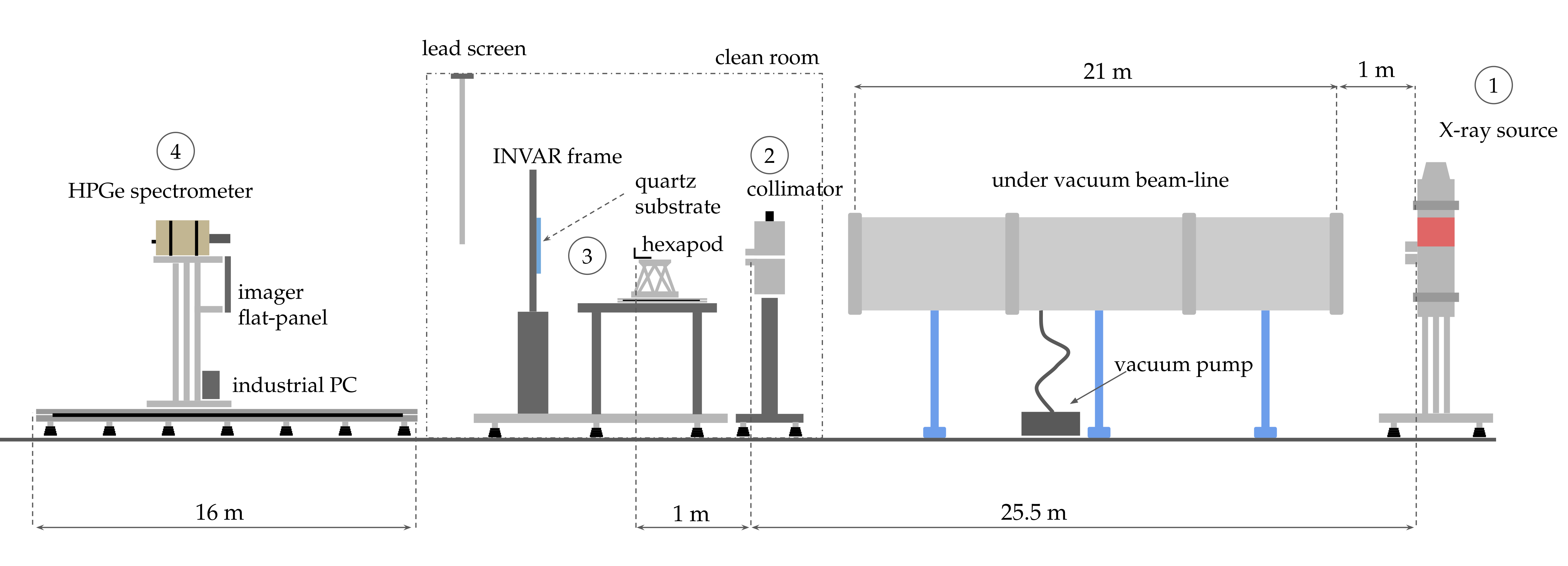}
    \includegraphics[scale = 0.40]{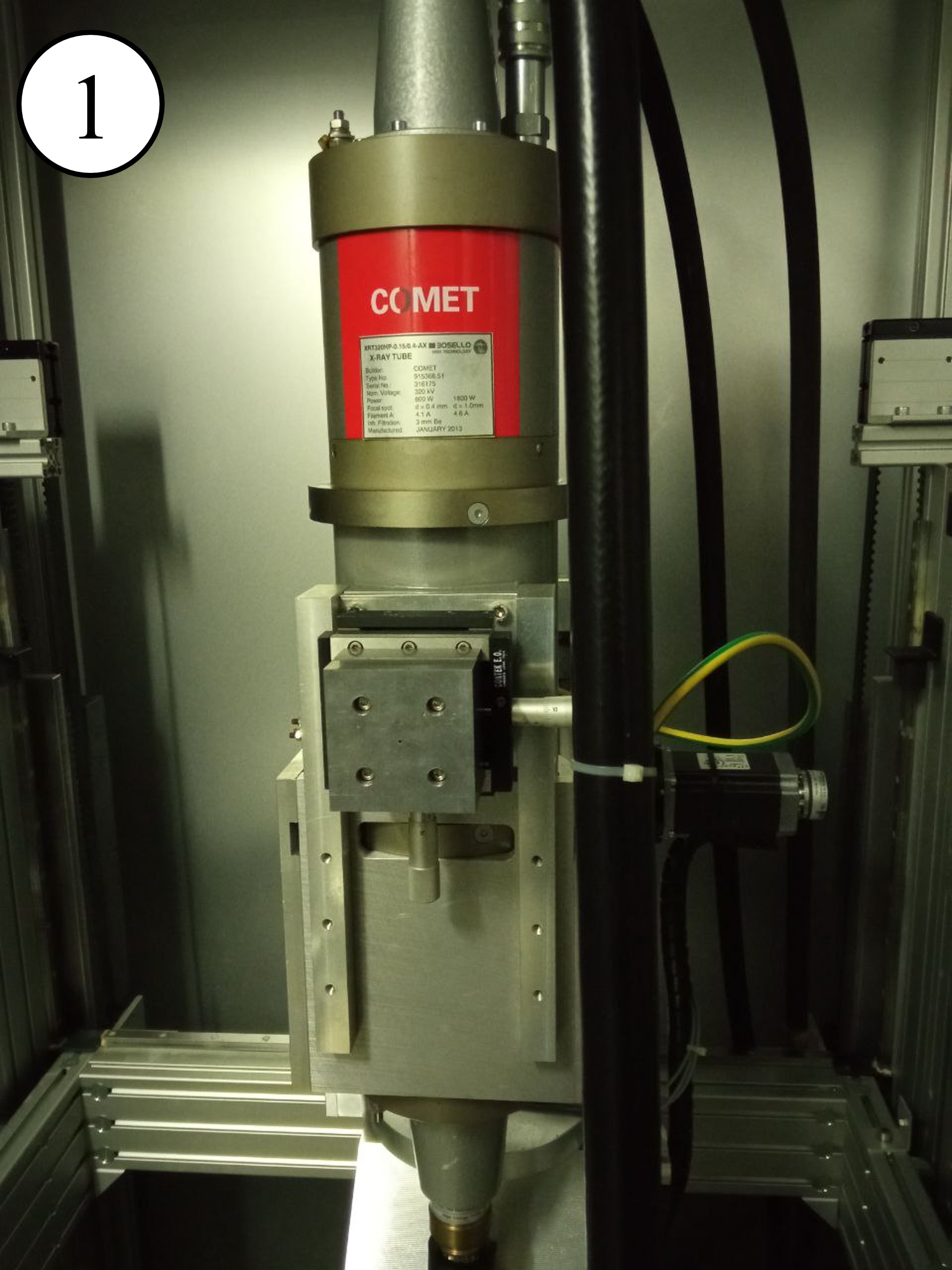}
    \includegraphics[scale = 0.398]{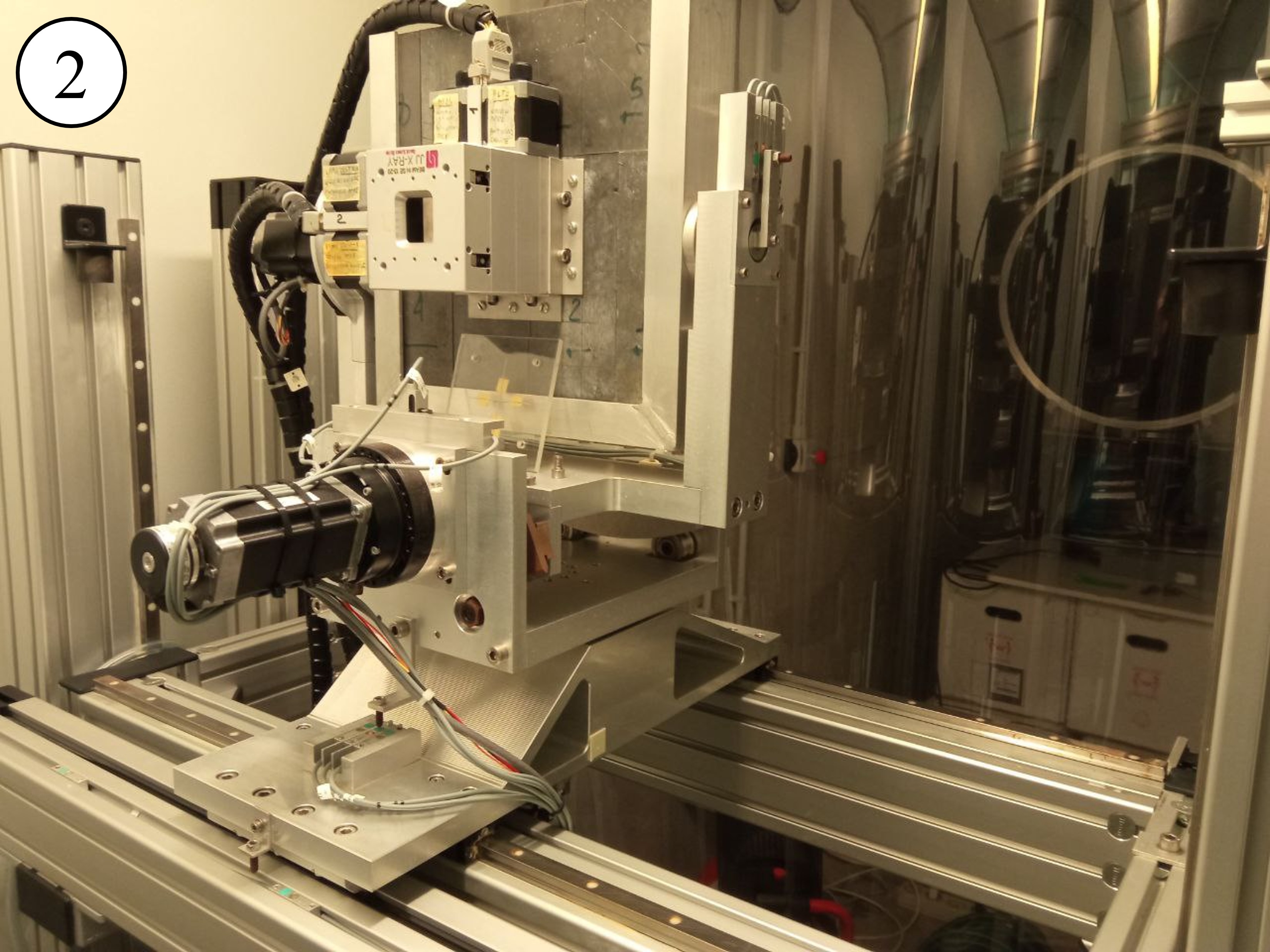}
    \includegraphics[scale = 0.28]{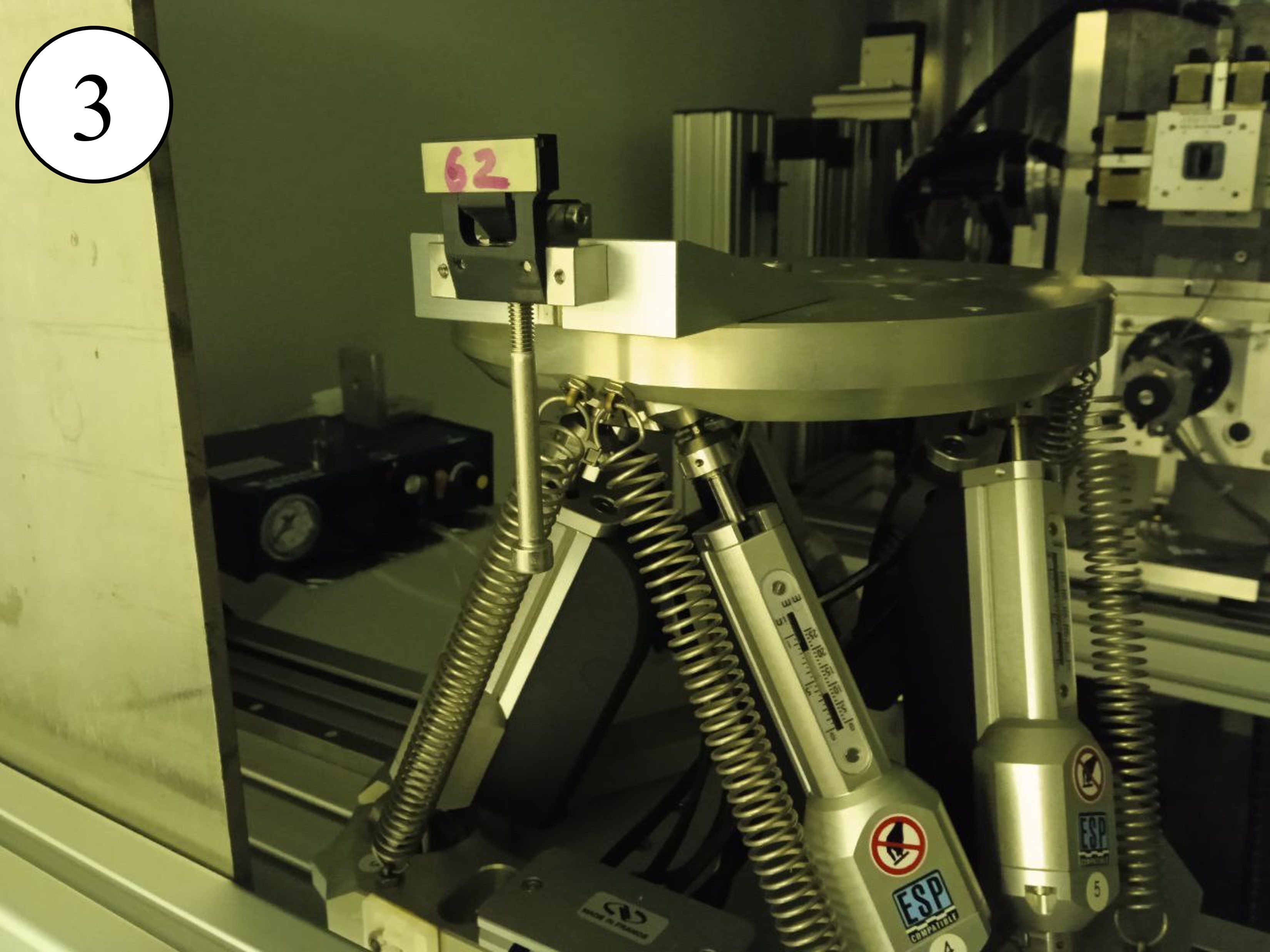}
    \includegraphics[scale = 0.64]{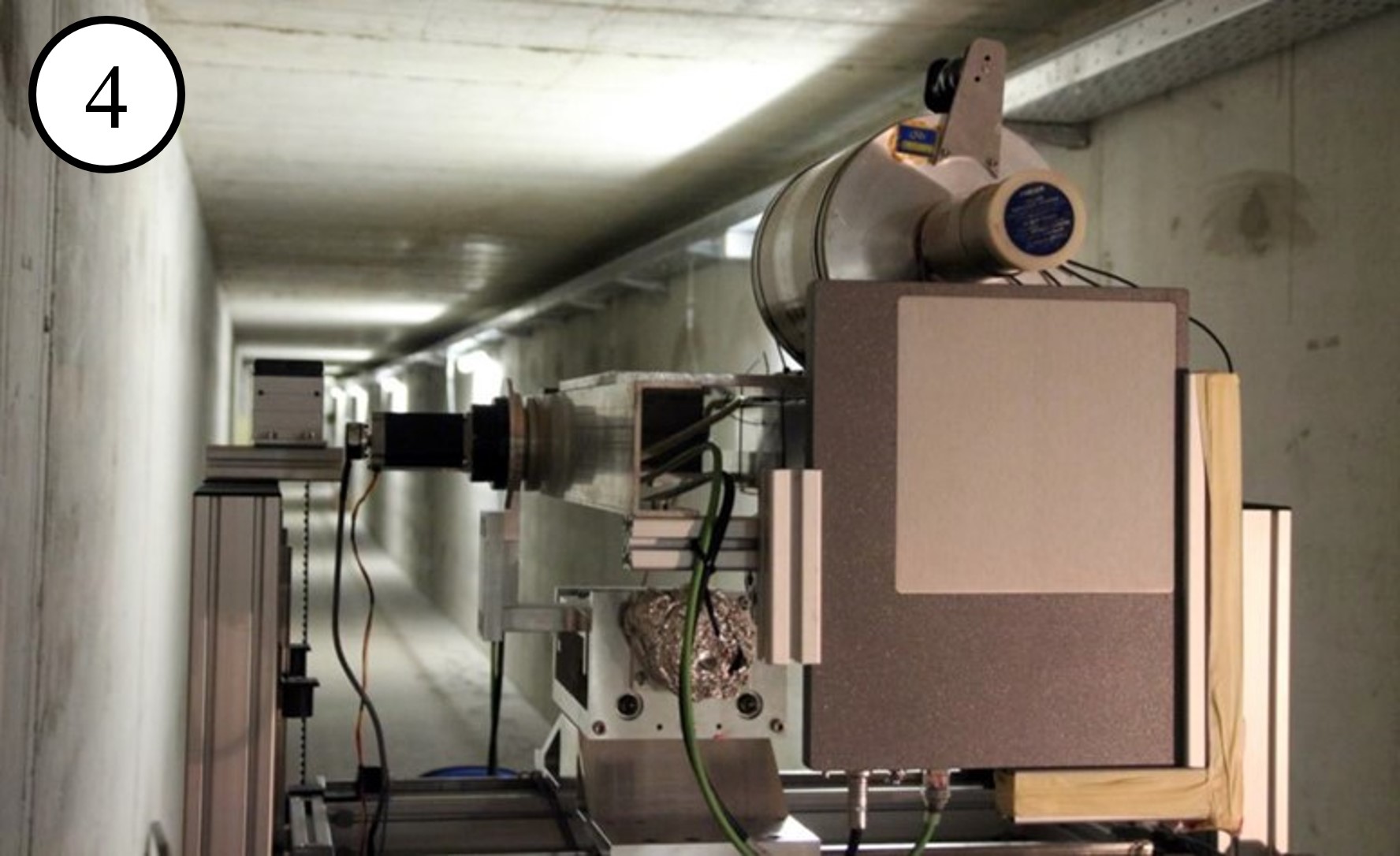}
    \caption{Top: Sketch (not at scale) of the LARIX-T facility at University of Ferrara, where the Laue lens module has been assembled and tested. Center left: the collimated X-ray source. Center right: the remotely controllable lead and tungsten collimator. Bottom left: the hexapod used to orient the crystals; the custom crystal holder is also visible with a mounted crystal. Bottom right: the suite of available detector, an HPGe spectrometer and a flat-panel imager Perkin-Elmer (200 $\mu$m spatial resolution).}
    \label{fig:set_up}
\end{figure}

The prototype was built in the 100~m long tunnel (LARIX-T) of the LARIX laboratory 
\footnote{https://larixfacility.unife.it/} 
of the University of Ferrara. 
A scheme of the facility is shown in Fig. \ref{fig:set_up}, top. The facility consists of a 26.5~m beamline working in the 50 -- 320 keV energy range.
The X-ray beam is produced by an X-ray tube equipped with a tungsten anode (Fig. \ref{fig:set_up}, top left) with a focal spot size of 0.4~mm. A 20~mm thick tungsten plate with a 3~mm diameter hole and a 50~mm thick lead shield with a 1~mm diameter hole are placed in front of the exit window to reduce the beam divergence; the two collimator plates are immediately in front of the X-ray tube exit window. The X-ray beam passes inside a 21~m long vacuum pipe and then through a motorized slit collimator with four independently motorized 20 mm thick tungsten blades (Fig.~\ref{fig:set_up}, top right). For this experiment, the collimator aperture was set in order to obtain a beam dimension of 10 $\times$ 10 mm$^2$. Given the distance of 26.5~m between X-ray source and quartz substrate and the beam size on the crystal, the divergence of the x-ray source over the crystal area is about 78~arcsec. This set-up is designed to reduce the divergence of the incident beam to approximate the illumination conditions from a source placed at infinite distance from the target\cite{Virgilli2014}. 

The crystals are positioned using a customized holder, which is mounted on a high precision 6-axis Hexapod HXP100-MECA from Newport (translation accuracy of 1 $\mathrm{\mu}$m and a rotation accuracy of $\mathrm{2 \times 10^{-5}~rad}$) (Fig. \ref{fig:set_up}, bottom left). The customized holder, visible in Fig. \ref{fig:set_up}, support the crystals from back, top side, and right side without exerting any force, while a small, movable bar support the crystals from the botton and can be adjusted in such a way that the crystal is clamped between top and bottom part of the holder.
The collimator and the hexapod are placed inside an ISO8 clean room. 
On the same carriage of the hexapod an INVAR steel frame is also installed. This frame hosts the quartz glass which is used as substrate for the module of the Laue lens under realization. Both the quartz glass and the INVAR steel have been chosen for their extremely low thermal expansion coefficients (0.55~ppm/$^o$C and 1.2~ppm/$^o$C, respectively).

A further carriage is placed on a movable rail, which allows the detector to be moved from a minimum distance from the lens frame of $\sim$ 8~m to a maximum distance of $\sim$ 23~m.  A Perkin Elmer Cesium Iodide (CsI(Tl)) digital X-ray flat panel detector with 1024 $\times$ 1024 pixels,  200 $\mu$m pixel size and sensitive in the broad energy range 20 keV – 15 MeV has been used to detect the diffracted signals by each crystal and for their mutual alignment (Fig. \ref{fig:set_up}, bottom right).

Given the divergence of the beam, the imager is placed at a distance $\mathrm{F_D}$ from the crystals, obtained by\cite{Virgilli2016}: 
\begin{equation}
    \frac{1}{F_D} = \frac{1}{D} + \frac{2}{R_c}
\end{equation}
Where $\mathrm{D}$ is the distance between the source and the sample holder and $\mathrm{R_c}$ is the curvature radius of the crystals. In the ideal case of a crystal with a curvature radius of 40.0 m, $\mathrm{F_D}$ is 11.4~m instead of 20.0~m. Note that X-rays get diffracted only once in Laue lens configuration, therefore the beam diverges contributes to reduces the focal lenght.

\subsection{Prototype Assembly}

\begin{figure}
    \centering
    \includegraphics[scale = 1.5]{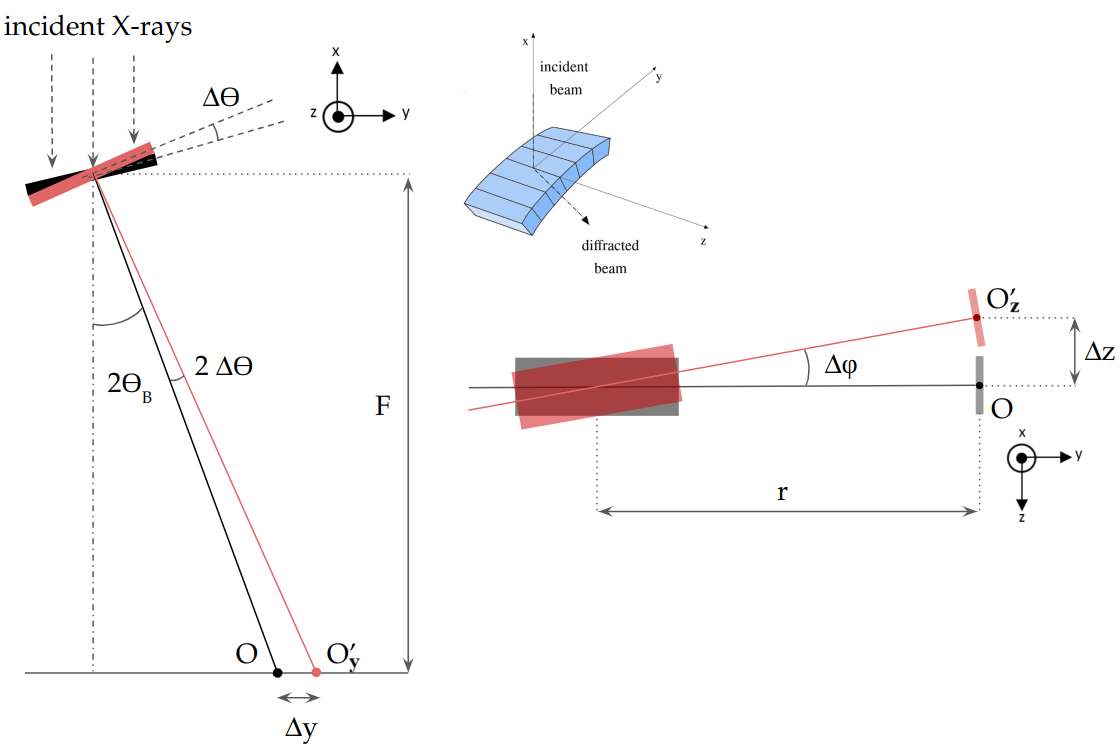}
    \caption{Scheme showing how the deviation of one crystal from the nominal positioning angles ($\theta$ and $\phi$) affect the position of the diffracted beam. Black points $O$ represent the nominal diffracted position. The variation 
    $\Delta$$\theta$ along the Bragg angle $\theta_B$ (left) shifts the diffracted signal in $O'_y$ by $\Delta {\rm y}=F\, \tan{(2\Delta\theta)}$. The variation $\Delta\phi$ of the polar angle respect to the nominal position results in a shift of the diffracted beam from $O$ to $O'_z$ by an amount $\Delta{\rm z}=r\, \tan{\Delta\phi}$ (right).}
    \label{fig:angles_expl}
\end{figure}

The main objective of this research is to fabricate a test module consisting of 12 bent germanium crystals. {With reference to Fig. 
\ref{fig:angles_expl}, for a parallel beam impinging on the lens,
the proper orientation of the diffracted beam towards the Laue lens focus is achieved by adjusting the crystal with respect to the $\theta$ angle, which is related to the Bragg angle, and to the polar, or radial, angle $\phi$. 
We aim to be able to build an astrophysical Laue lens with a PSF of the order of 30 arcsec HPD in all its energy band, which mean that we require to position the Bragg's angle of the crystal with an accuracy $<$10 arcsec\cite{Frontera2021}. The misplacement of the image of a crystal due to a wrong incident angle scales with the focal of the lens, so, with a long focal such as our case, it is very important to keep this type of misalignment as small as possible. On the polar angle, instead, we can have a less strict requirement of an accuracy of at least 5 arcmin, since the misplacement of the image induced by a polar angles misalignments scales with the radius of the lens, which is about one order of magnitude smaller than the focal length.}
Even with a limited number of crystals, such a cluster is representative of the elemental module that constitutes a complete Laue lens. 
The position of each crystal is set under the control of the X-ray beam to align the positions of the centroid of their images all on the same point on the detector. Each crystal is bonded to the substrate through the following procedure:
\begin{enumerate}
\item the crystal is mounted on the hexapod and the collimator is positioned in such a way that the central part of the crystal is illuminated;
\item the position of the diffracted signal is measured with the flat panel. From the difference between the nominal and measured positions of the diffracted signal, the Bragg and polar angles are evaluated and provided to the hexapod in order to correctly orient the crystal;
\item a drop of glue is dispensed on the substrate in correspondence to the central part of the crystal. The polished side of the crystal is then pressed in contact with the adhesive. {We chose to apply the adhesive on the polished side to avoid that the glue induces an excessive stress on the crystals, possibly changing their curvature radii.} The typical distance between crystal and substrate is about 150-200~$\mu$m. A final check of the position of the diffracted signal is made. By fitting the diffraction profile in both directions, the measurable positional accuracy is better than 0.5 pixels, i.e. 
of the order of 1.5 - 2 arcsec.
\item the glue is cured from the back of the quartz substrate by using a UV lamp DYMAX BlueWave 75 with the use of a light-guide. We alternated between short light shots and longer dark times. The idea below this procedure is that the light shots partially cure the glue, so during the dark times there is still room to correct the position of the crystal in case that the expected shrinkage of the glue moved the crystals from its desired position. For each crystal, we performed 60 cycles of 0.5~s of light and 45~s of dark, then 20 cycles of 2~s of light and 45~s of dark. The total light time given to each crystal is 70~s, while the total dark time is 3600~s, so the curing process of each crystal takes 1 hour. 
\item the crystal is released from the hexapod and the position of the centroid of its diffracted image is measured. The release of the crystal is typically the  most crucial phase of the bonding process, since the combined effects of the adhesive force of the glue, gravity, and any backlash from the release of the crystal clamp can drastically change the position of the crystal if the bonding process was not performed properly. The position of the diffracted signal for each crystal is monitored at regular time intervals, with a particular interest in evaluating its long term stability.
\end{enumerate}

\section{Data analysis and results}
\label{analysis}

\begin{figure}
    \centering
    \includegraphics[scale = 0.2]{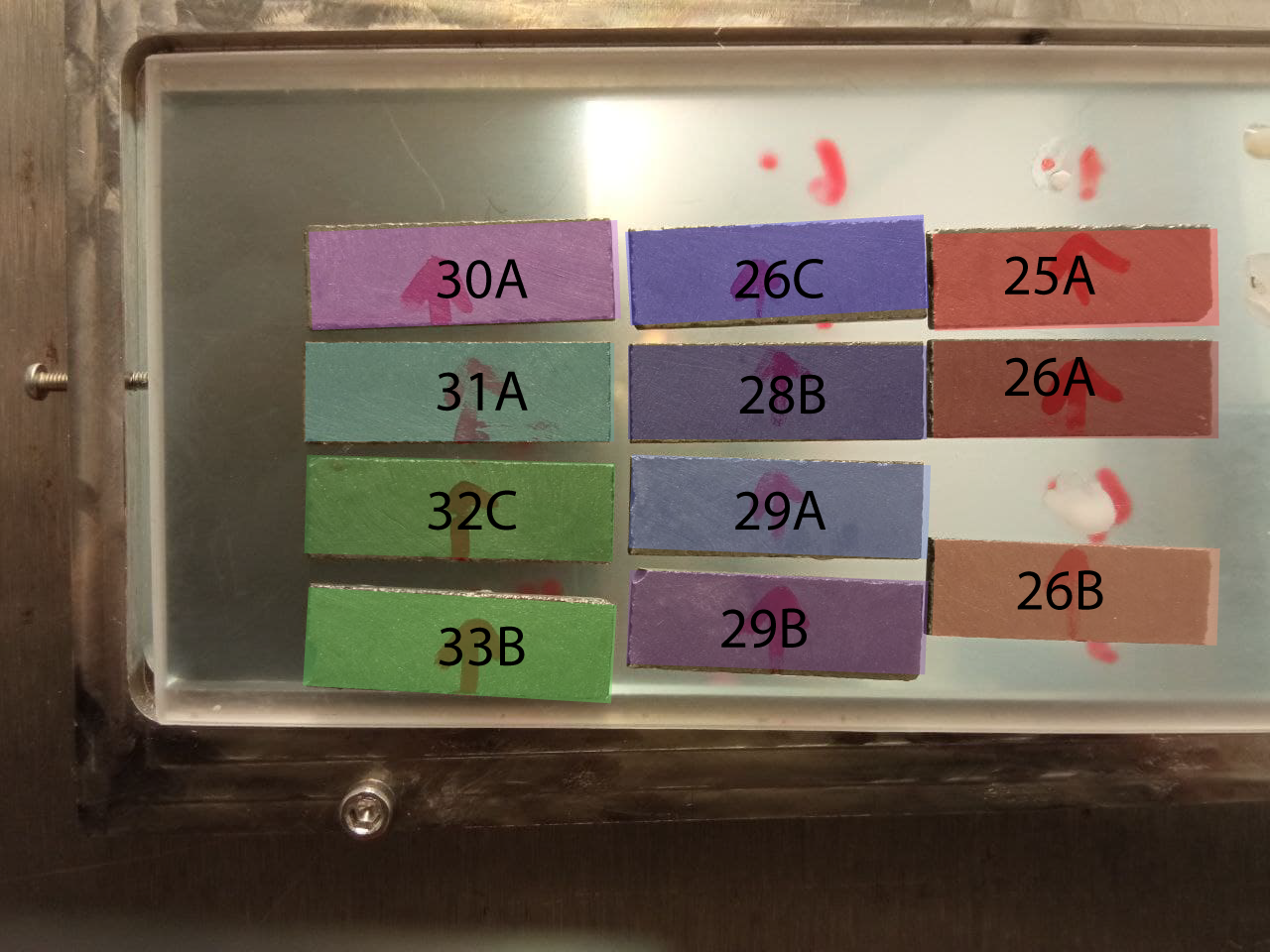}
    \includegraphics[scale = 0.135]{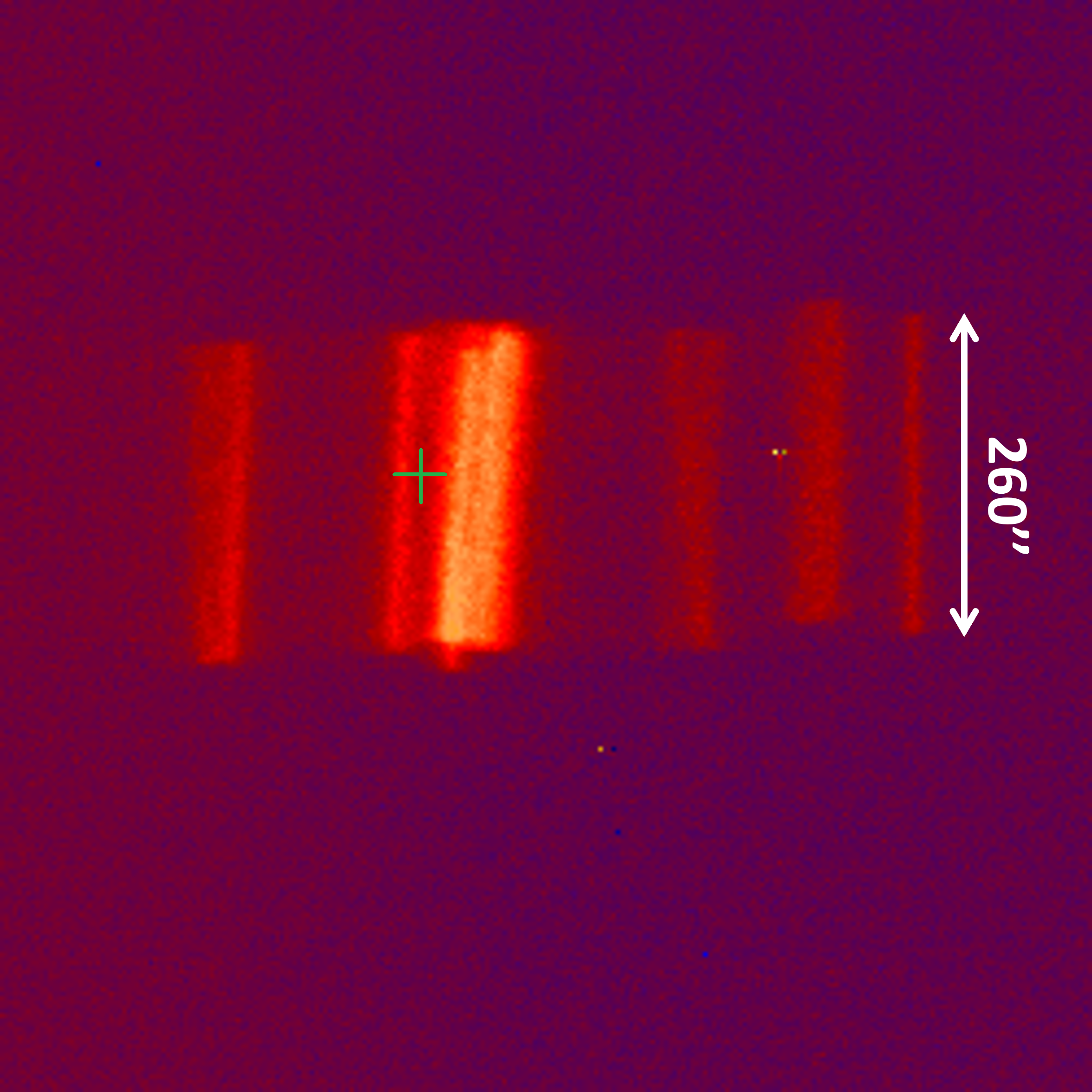}
    \caption{Left: The eleven crystals bonded on the quartz substrate fixed on the INVAR support. The color overlapped on each crystal corresponds to the colors used to distinguish each crystals on the plots in Fig. \ref{fig:misal_distrib}. Right: diffracted image produced by the 11 crystals fixed on the substrate and illuminated simultaneously. The center of the green cross was the target position on which we tried to align the image of each crystal.}
    \label{fig:module_assembled}
\end{figure}

\begin{figure}
    \centering
    \includegraphics[scale = 0.30]{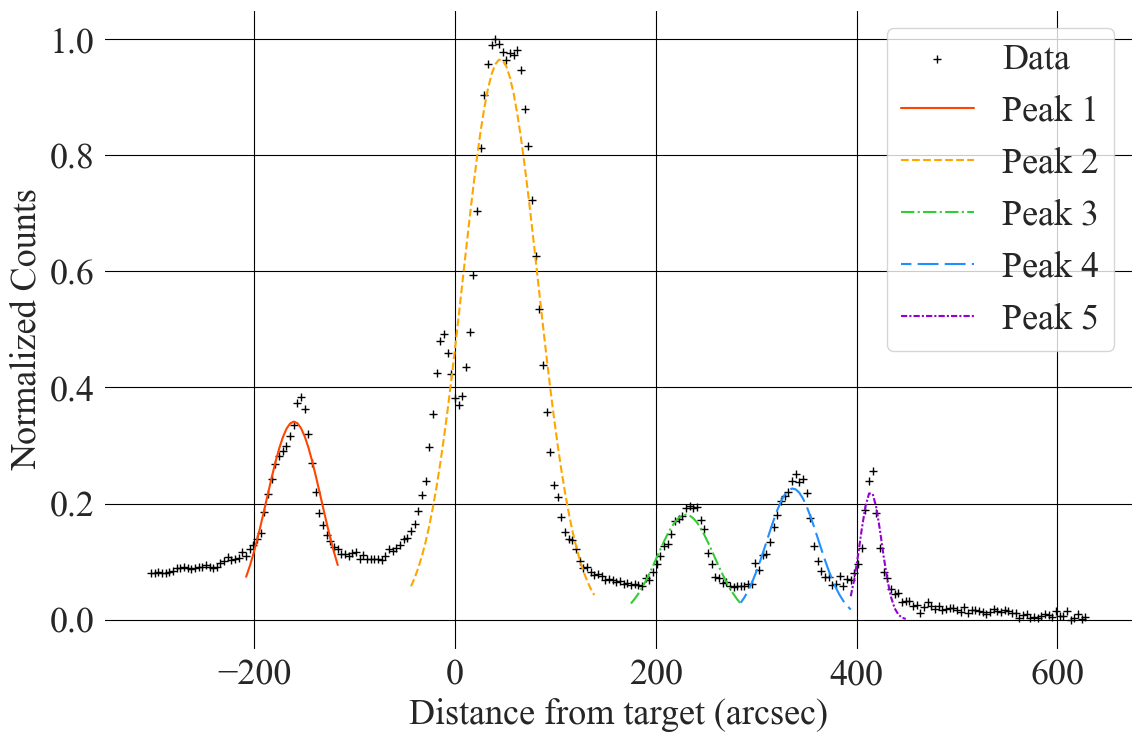}
    \includegraphics[scale = 0.30]{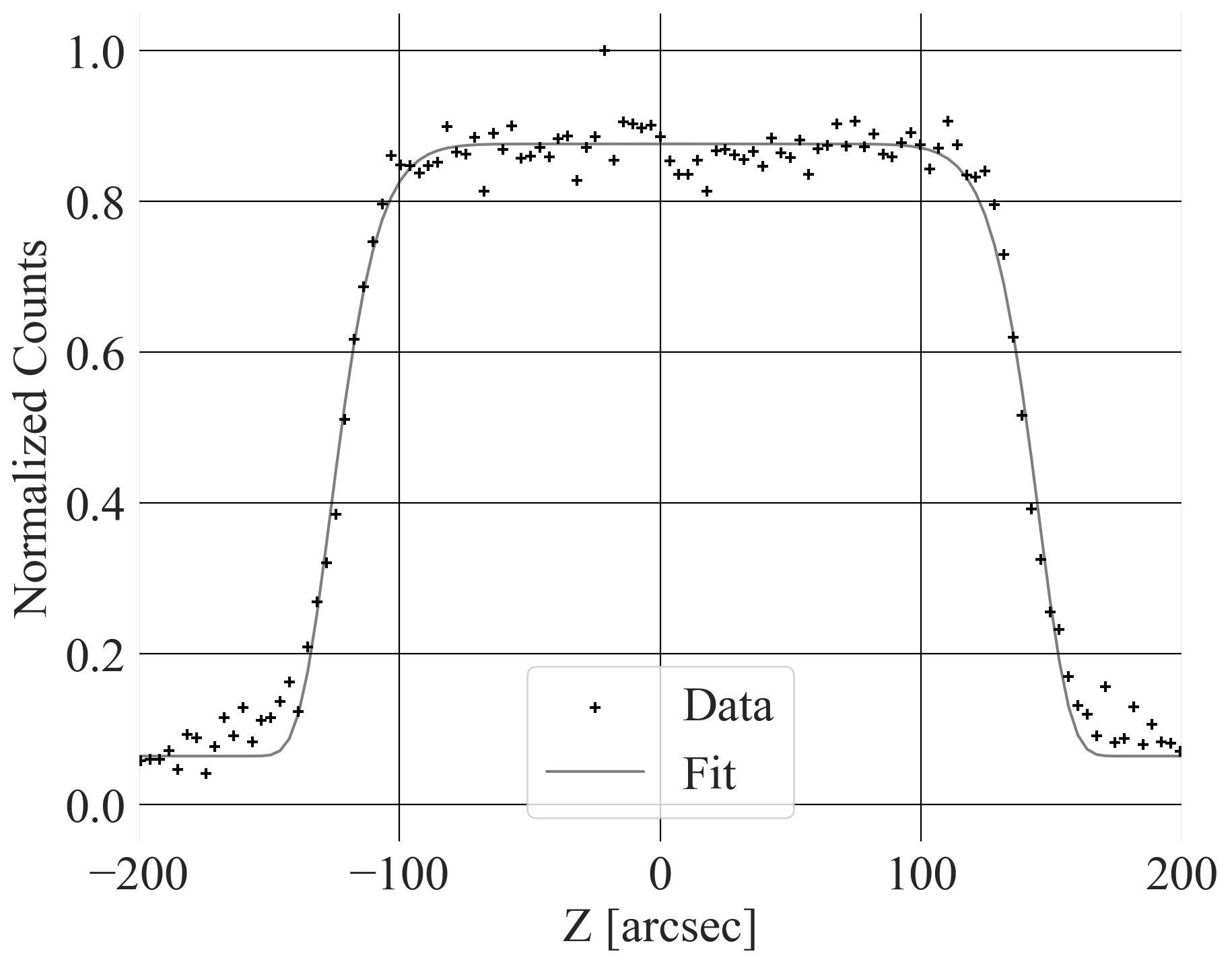}
    \caption{Left: Profile along the focusing direction of the image produced by the assembled Laue lens module. The main peak includes the overlapped images of 7 crystals. Right: Profile along the non-focusing direction of the image produced by the assembled  of the image produced by the assembled Laue lens module.}
    \label{fig:11_xtal_images}
\end{figure}

\begin{figure}
    \centering
    \includegraphics[scale = 0.3]{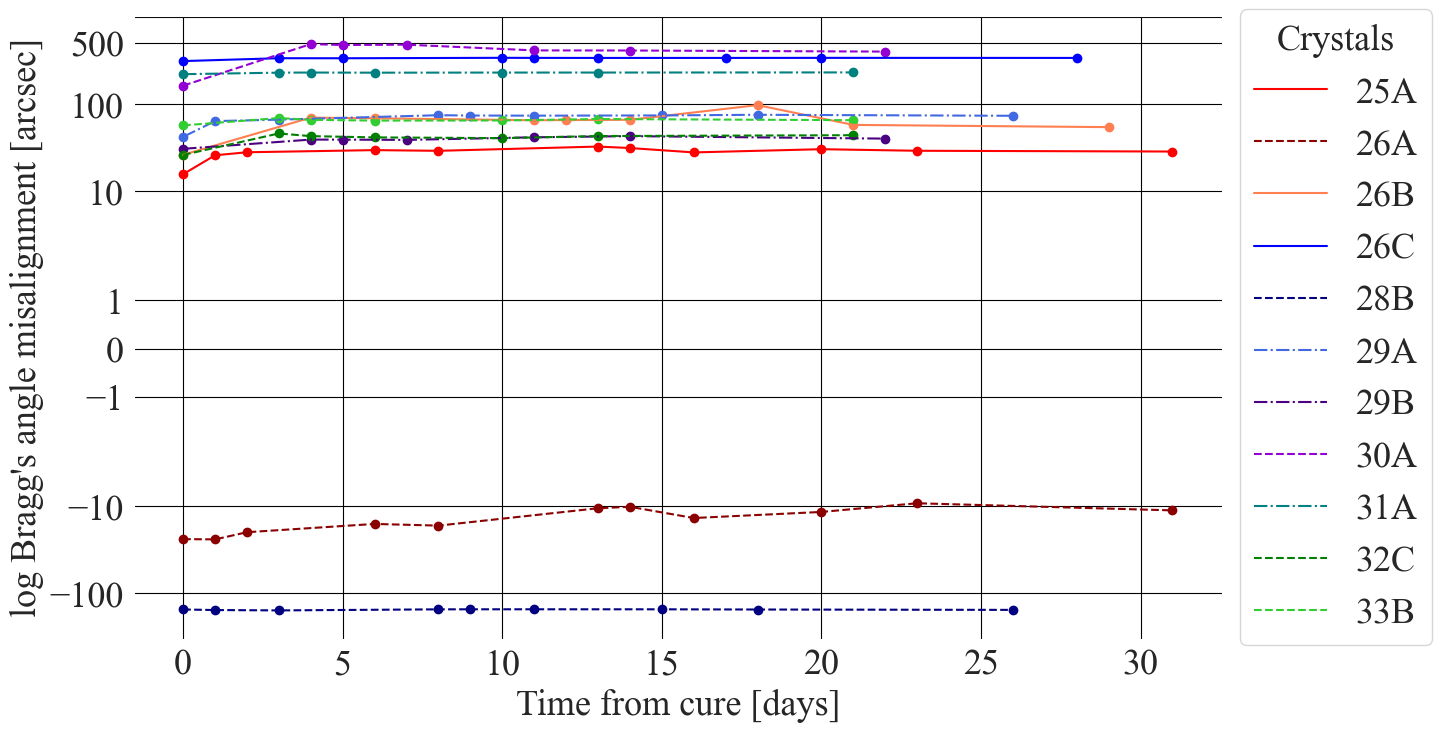}
    \includegraphics[scale = 0.3]{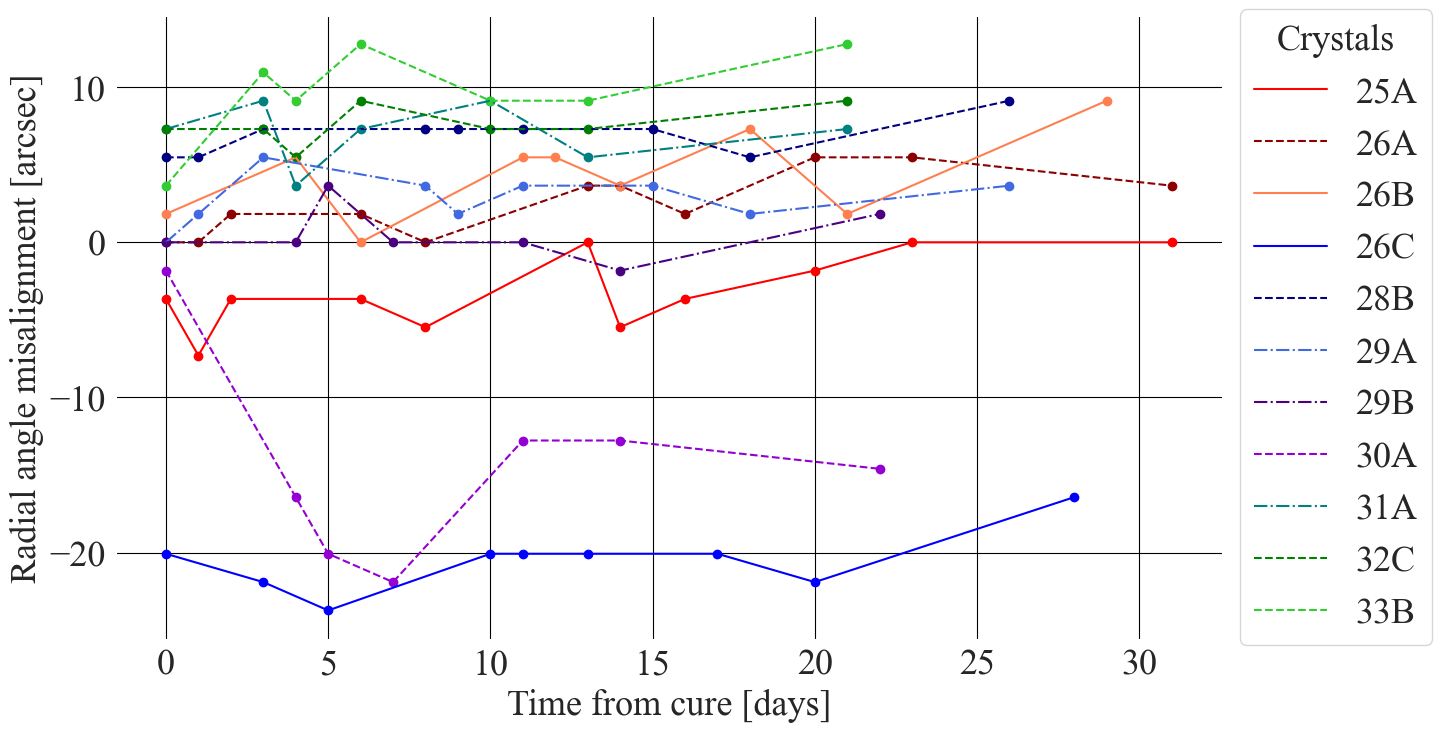}
    \caption{Measured misalignment in the Bragg's angle (top) and polar angle (bottom) for every crystal as a function of time after bonding.}
    \label{fig:misal_distrib}
\end{figure}

The assembled module of Laue lens is shown in Fig.~\ref{fig:module_assembled} (left), while the image of the crystals illuminated together is shown in Fig.~\ref{fig:module_assembled} (right). The bonding procedure is quite time-consuming, since continuous checks are needed on the position of the crystals throughout the process. Of the ten days passed from the bonding of the first to the last crystals, seven were working days, resulting in the bonding of 1-2 crystals a day. To check the positional stability of the crystals in the assembled module, we repeated the measurement of the position of the diffracted beam from each crystal daily, for 31 (21) days after the assembly of the first (last) crystal. After this period, we noted that the crystals' position remained stable for at least one week, when we stopped the daily measurements. Very preliminary results of our assembled module were reported in a previous work\cite{Ferro2022}. The final sector was made by 11 crystals instead of 12 because the crystal 26B got detached from its original position due to an assembly mishap and left a white spot of cured glue on the quartz substrate (clearly visible in Fig.~\ref{fig:module_assembled}), which rendered it impossible to bond another crystal on the same position without risks of damage to the assembly.
Finally, the profile of the image along the focusing direction is shown in Fig.~\ref{fig:11_xtal_images},~left and the profile along the non-focusing direction is shown in Fig.~\ref{fig:11_xtal_images},~right. The images shown in Fig.~\ref{fig:module_assembled} have been taken 31 days after the bonding of the first crystal. Seven of the 11 crystals are aligned in the central peak of the image, while four crystals (30A, 26C, 31A, 28B) are strongly misaligned along the focusing direction. We can see that the width of the diffracted peaks produced by the crystals 26C, 31A, 28B are twice the width of the peak generated by crystal 30A. We believe that this effect can be ascribed to the stress generated by the glue on the crystals, which in some cases can be strong enough to deform their curvature radius.
In the focusing direction, the combined image of the crystals show five distinct peaks: the major peak results from the combination of seven well-aligned crystals, is centered at a distance of 45 arcsec from the expected position (see Fig.~\ref{fig:11_xtal_images}) and has a FWHM of 88~arcsec. The four smaller peaks are generated by the outlier crystals. Position and FWHM of every peak are reported in Table~\ref{tab:real_peak_fits}. On the non-focusing direction, the profile of the image can be fitted by a box shape centered on $\mathrm{9.2 \pm 0.3}$~arcsec from the target position and with a width of $\mathrm{267 \pm 7}$ arcsec, which corresponds to a linear width on the detector of 15 mm at the distance of 11.4~m, compatible with the expected crystal footprint taking into account the divergence of the X-ray beam.
The time stability of the assembly has been evaluated by measuring the position of the centroid of every crystal day by day after the cure. Time stability measurements on both the Bragg and polar angle can be seen in Fig. \ref{fig:misal_distrib}. After 31 days from the cure of the first crystal, the average misalignment value in the Bragg's angle is 100 $\mathrm{\pm}$ 50~arcsec, while the average misalignment value in the polar angle is 2 $\mathrm{\pm}$ 3~arcsec. The amplitude of the Bragg's angle distribution is 9.58 $\mathrm{\pm}$ 0.03 arcmin, while the amplitude of the polar angle distribution is  30 $\mathrm{\pm}$ 7 arcsec.
Taking into account that, the angle with respect to the incoming direction of the photons is twice the angle between the diffraction planes and the beam\cite{frontera2011} in transmission configuration, the assembly misalignment of the crystals are then obtained by dividing the angular misalignment we measured by two. This means that the average Bragg's angles' assembly misalignment is 50 $\mathrm{\pm}$ 25~arcsec and the width of the distribution is 4.79 $\mathrm{\pm}$ 0.02 arcmin, while the average polar angles' assembly misalignment is 1.0 $\mathrm{\pm}$ 1.5 arcsec.

\begin{table}
    \centering
        \caption{Fit parameters of the five peaks forming the combined image of the 11 crystals bonded on glass, projected along the focusing direction. The peaks are numbered in increasing order, from left to right.}
    \begin{tabular}{cccc}
    \\
    \toprule 
         Peak Number & Normalization & Mean [arcsec] & FWHM [arcsec]\\
    \midrule 
         1 & $\mathrm{0.341 \pm 0.009}$ & $\mathrm{-160.5 \pm 0.9}$ & $\mathrm{64 \pm 2}$ \\
        2 & $\mathrm{0.96 \pm 0.03}$ & $\mathrm{45 \pm 1}$ & $\mathrm{88 \pm 3}$ \\
        3 & $\mathrm{0.181 \pm 0.006}$ & $\mathrm{230 \pm 1}$ & $\mathrm{67 \pm 3}$ \\
        4 & $\mathrm{0.226 \pm 0.009}$ & $\mathrm{336 \pm 1}$ & $\mathrm{60 \pm 3}$ \\
        5 & $\mathrm{0.22 \pm 0.01}$ & $\mathrm{414.2\pm 0.8}$ & $\mathrm{25 \pm 2}$ \\
    \bottomrule
    \end{tabular}
    \label{tab:real_peak_fits}
\end{table}

\newpage
\section{Simulations}
\label{simulations}

The mean and width of the distribution of misalignment angles and the uncertainty of the radius of curvature were used to simulate both a sector and a whole Laue lens.
For this purpose, the {\it Laue Lens Library}~(\texttt{LLL}), which has been already described elsewhere\cite{Virgilli17}, was used. The \texttt{LLL} is a hybrid analytical + ray-tracing Monte Carlo tool developed to simulate different configurations of Laue lenses and evaluate their performance. At present, the Laue simulation tools assumes that the curvature radius of the crystals, as well as the angles $\theta$ and $\phi$, can be distributed according to a uniform or Gaussian profile with respect to an average value. However, the model does not include the effect of the variation of the curvature radius which may occur during the curing process due to the aforementioned stresses.

The simulated mosaicity of the crystals is set to $\mathrm{10}$~arcsec, as required for the batch of crystals used in the experimental campaign.

\subsection{Module of a Laue lens: laboratory configuration}
First, a Laue lens module made from bent Ge (220) was simulated, assuming a  uniform distribution of misalignment errors on both Bragg's and polar angles of the crystals. The mean and width of these distributions are the same as measured from the assembled module.
In addition, we included in the simulation a uniform random distribution of the curvature radius of the crystal, with mean value of 39.7~m and width of 1.0~m. These values were obtained from the measurements of the curvature radius of a sample of 82 bent crystals of Ge(220) specifically prepared for this project and measured at CNR-IMEM\cite{Ferro2022}~. The set of 11 crystals used for this demonstration prototype is part of the same batch of samples. Assembly misalignment angles and curvature radius of each crystal were sampled from those random distributions.

In the simulations, we used the geometrical configuration of source--lens prototype--detector system in our experimental setup, so we simulated a source-lens distance of 26.5 m, lens-detector distance of 11.4 m and source size of 0.4 mm.

\begin{figure}
    \centering
    \includegraphics[scale = 0.3]{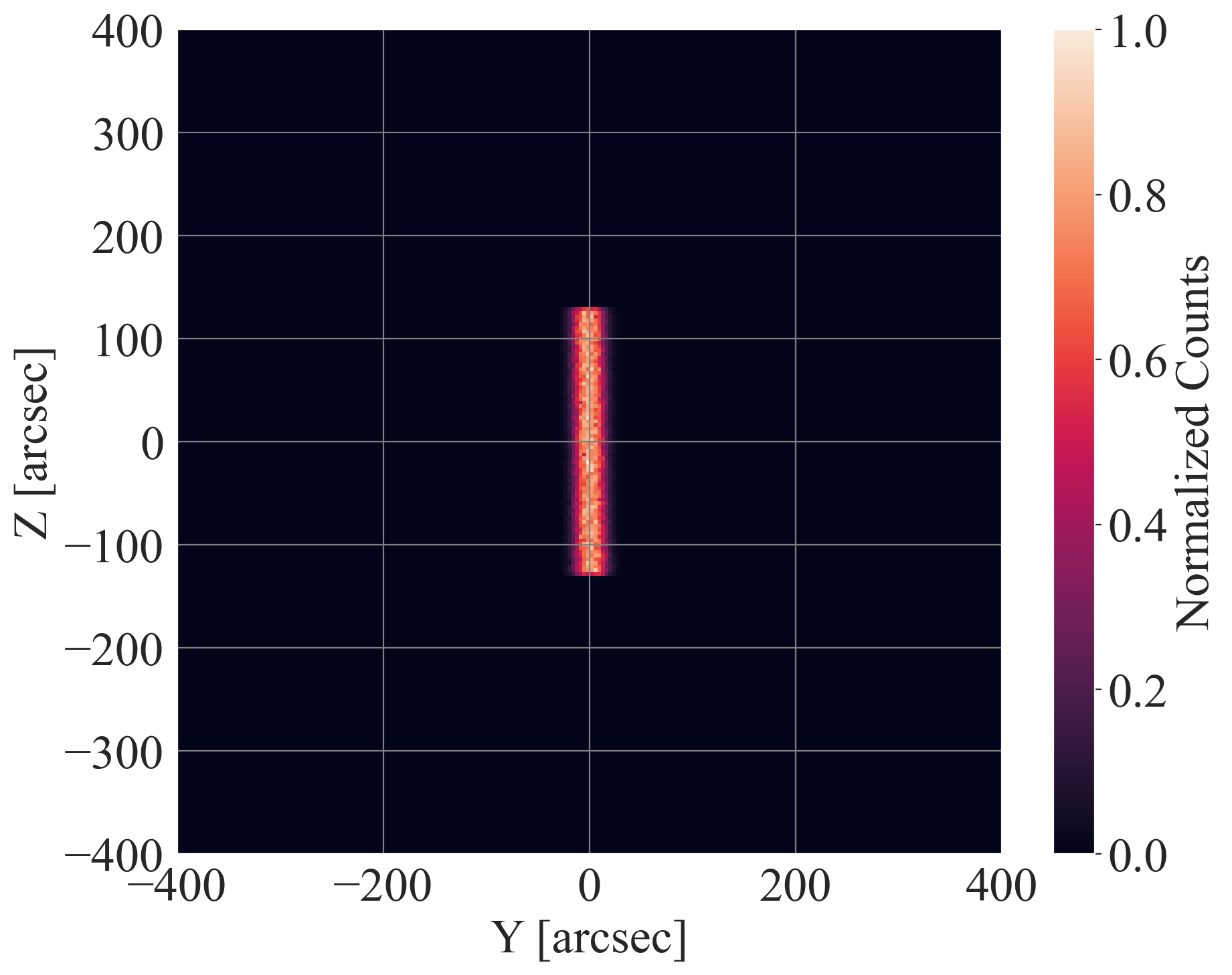}
    \includegraphics[scale = 0.3]{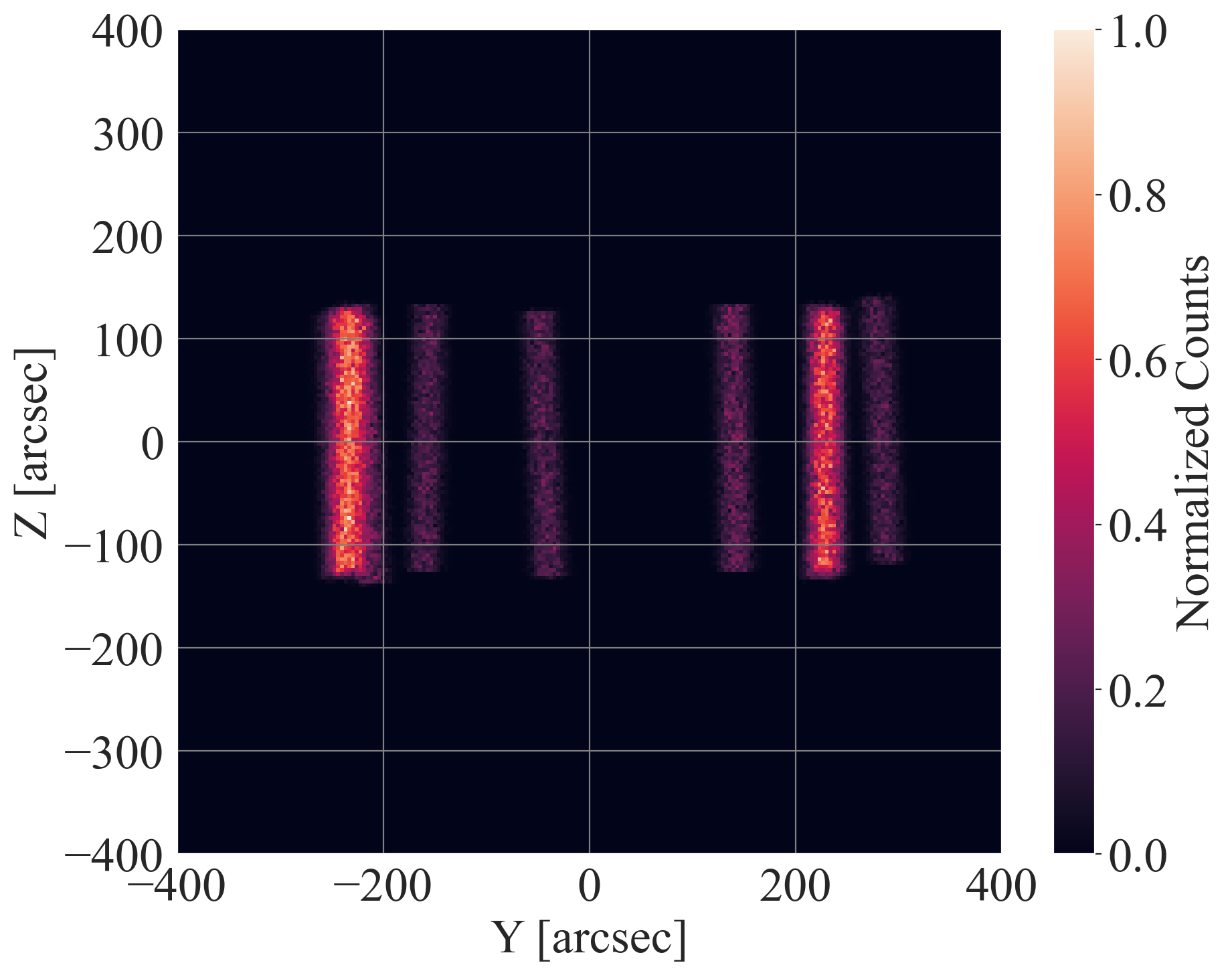}
    \includegraphics[scale = 0.3]{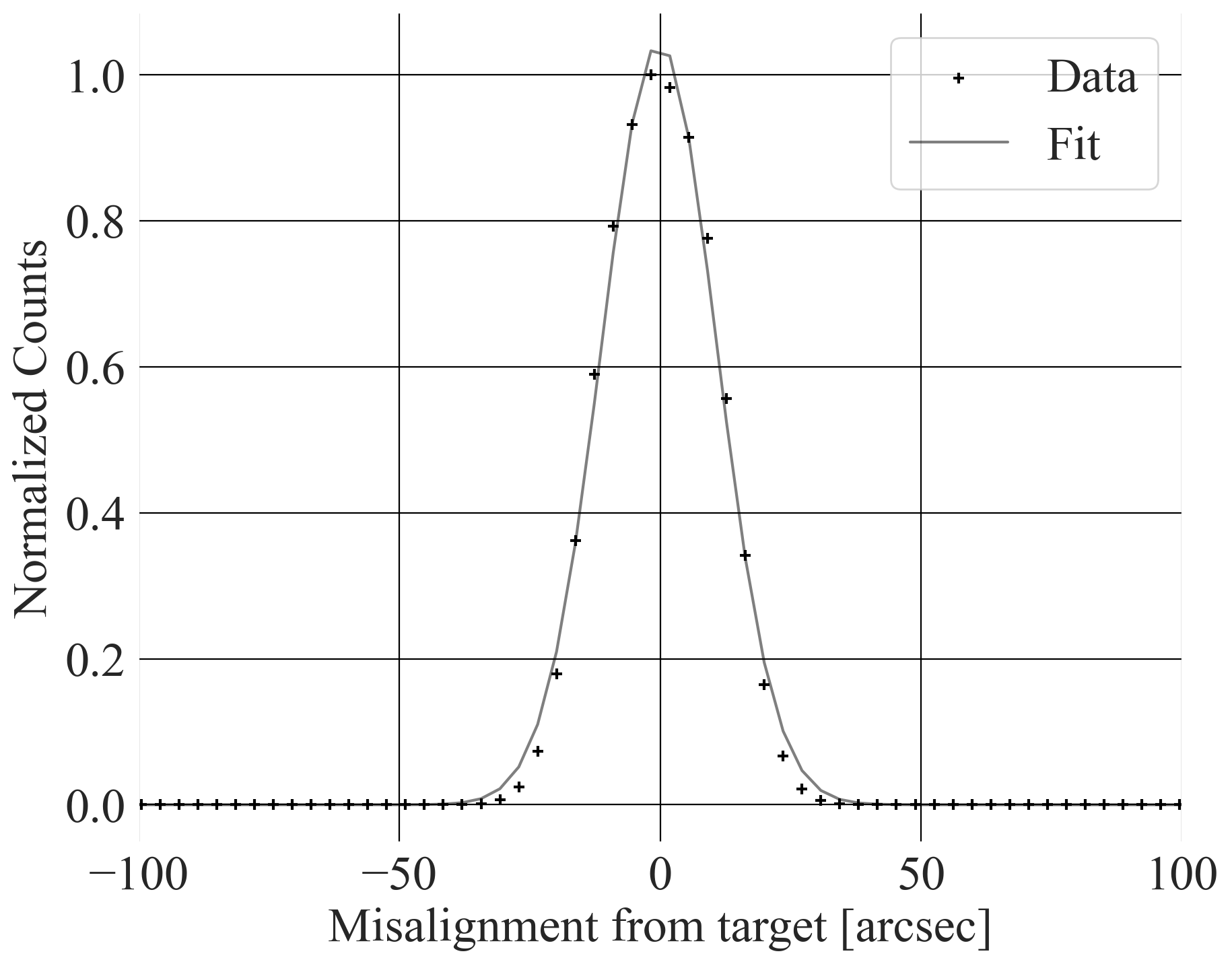}
    \includegraphics[scale = 0.3]{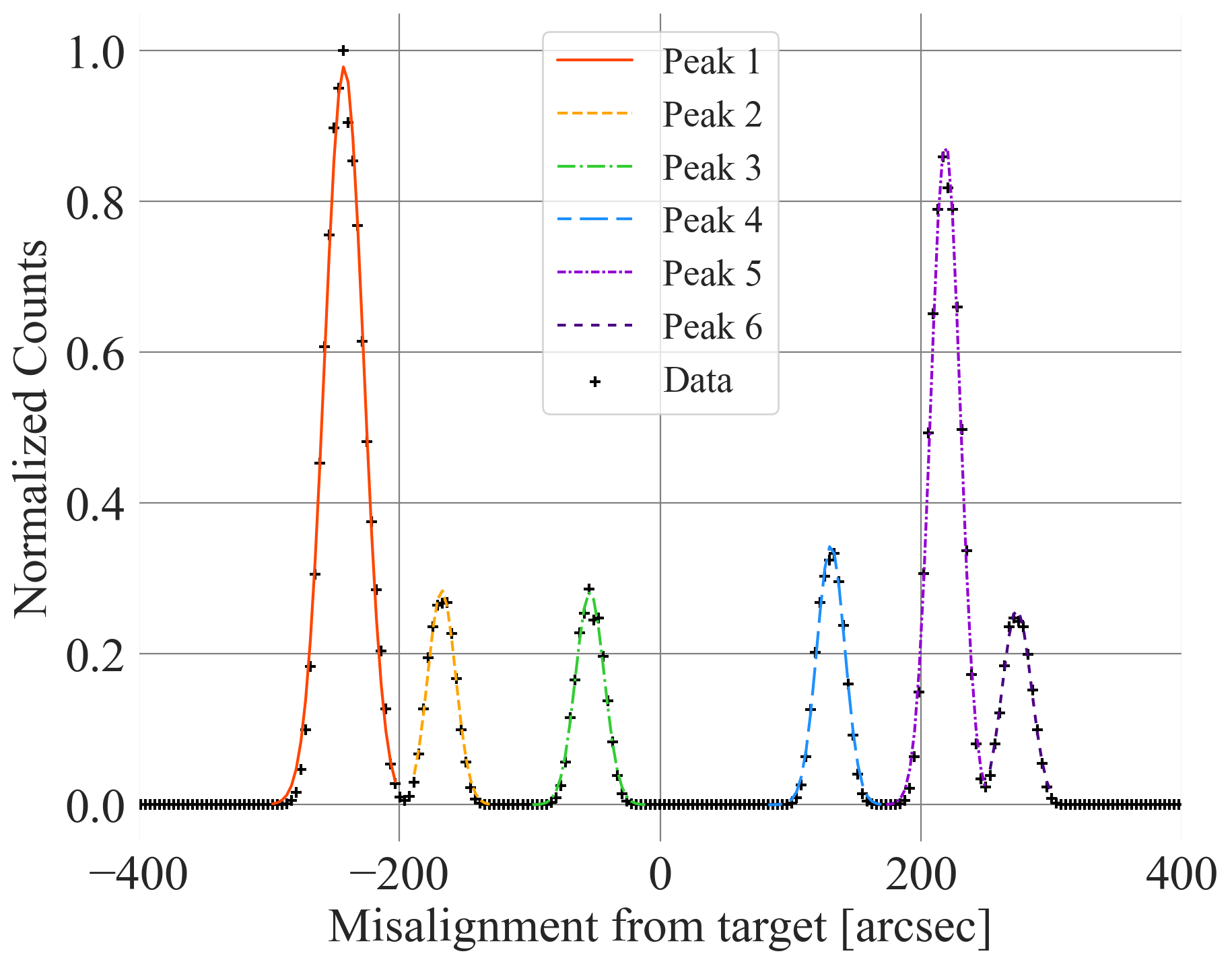}
    \includegraphics[scale = 0.3]{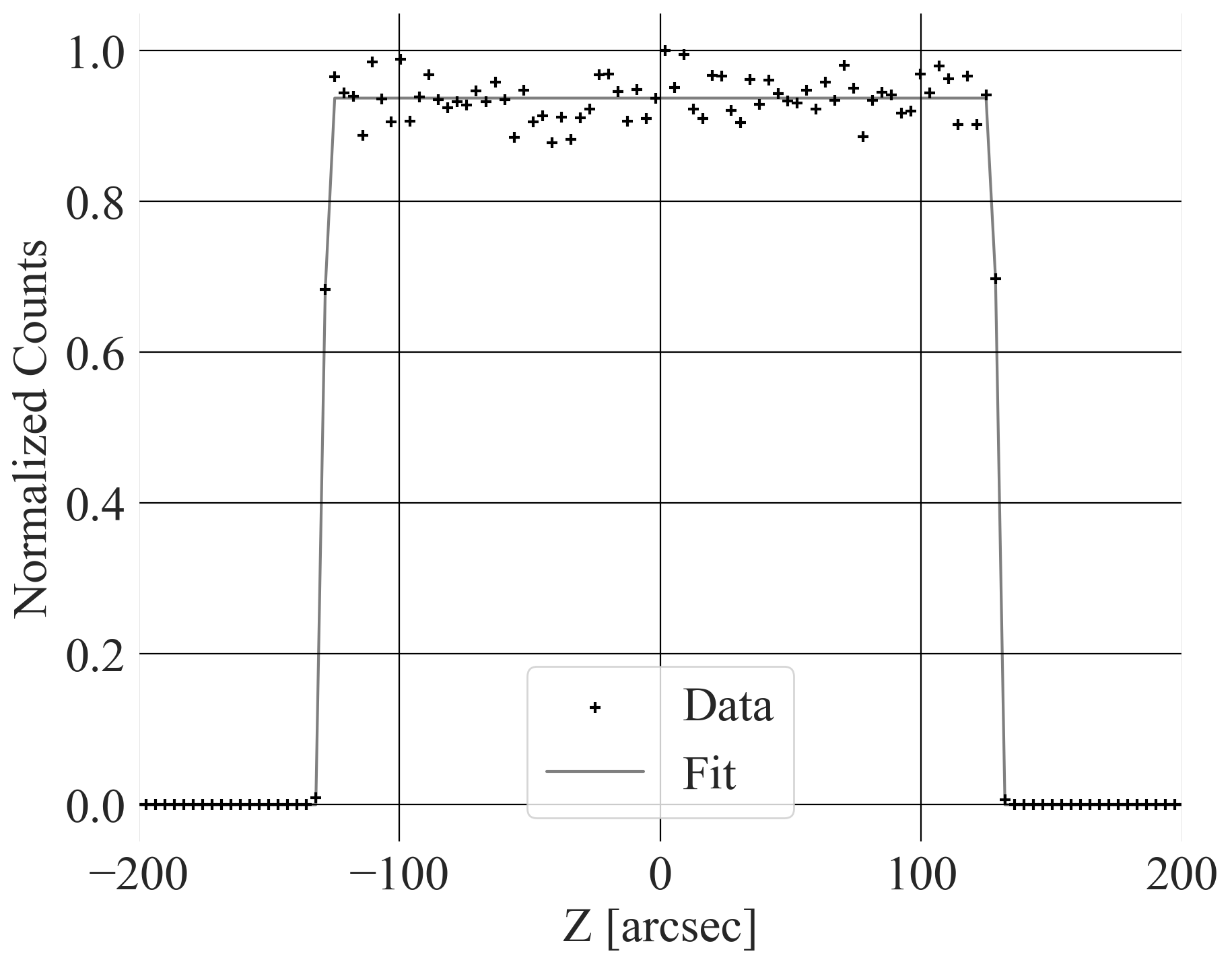}
    \includegraphics[scale = 0.3]{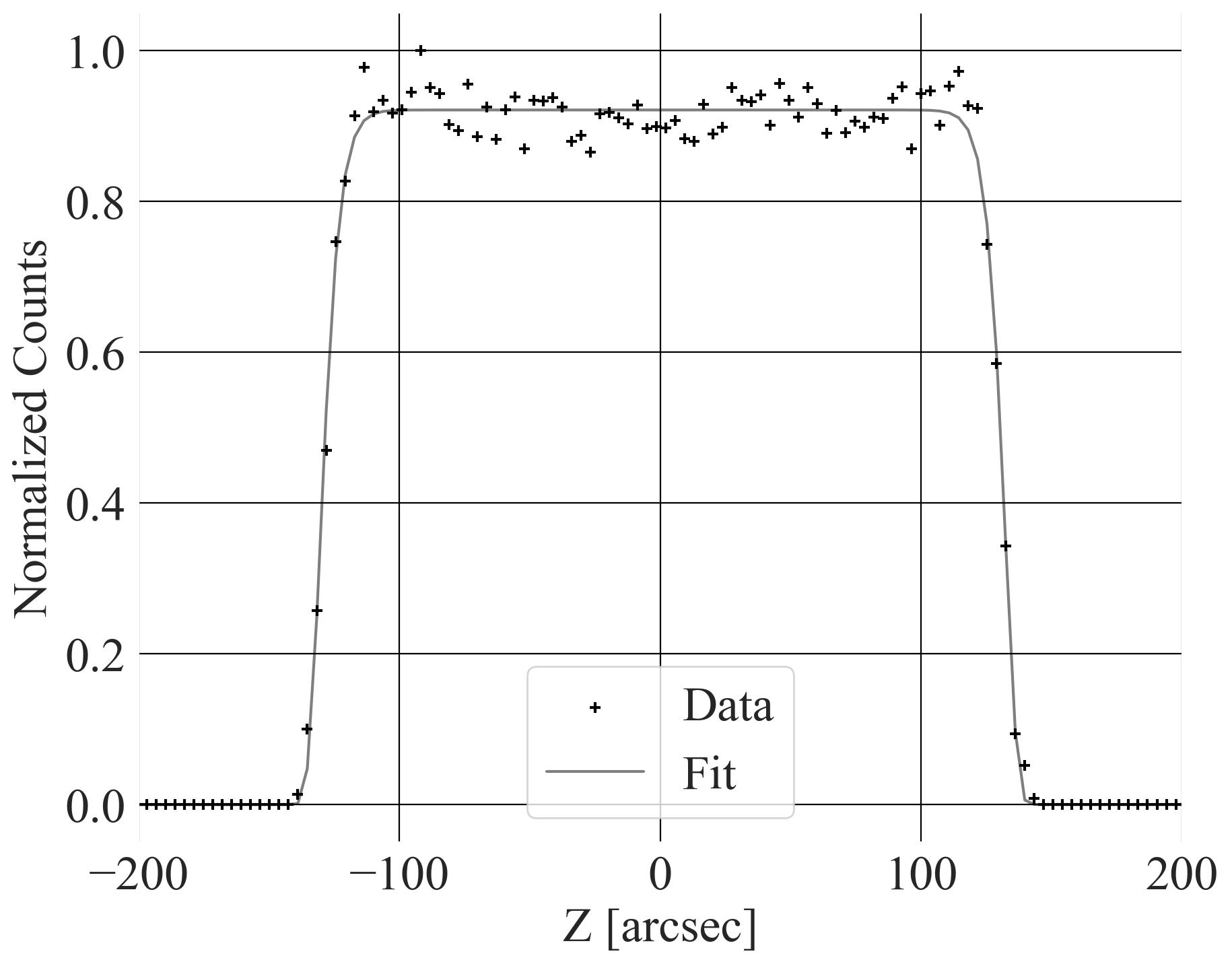}
    \caption{Comparison of the simulated images and profiles in the case of ideal crystal alignment configuration (left) and a configuration reproducing the laboratory setup with the observed alignment error  (right). Top: Focal plane images of an ideal and a real sector. Center: Profiles along the focusing direction. Bottom: Profiles along the non-focusing direction.}
    \label{fig:simulations_sector_lab}
\end{figure}

We also simulated the performance of a sector consisting of 12 perfect crystals of Ge(220), all bent with a radius of curvature of 40~m and with no misalignment error from the ideal position. This nominal module is taken as a reference for comparison with the experimental module.
Along the focusing direction (Y direction), the profile of the combined image from the 12 crystals has a Gaussian profile with a FWHM of $\mathrm{36.6 \pm 0.1}$~arcsec. 
Along the non-focusing direction (Z direction), due to the cylindrical curvature of the sample, no focalization is expected. The data are fitted with a box profile centered on $\mathrm{0.01 \pm 0.02}$~arcsec with width of $\mathrm{257 \pm 7}$~arcsec. The image and the profiles along the Z and Y directions of the ideal configuration sector are shown in Fig.~\ref{fig:simulations_sector_lab}, left.

Images and profiles along Z and Y direction for the distorted/misaligned configurations are shown in Fig.~\ref{fig:simulations_sector_lab}, right.
In the latter condition, the overall image of the crystals projected along the focusing direction shows six separate peaks which can be fitted with Gaussian profiles. The parameters used to fit the peaks are reported in Tab.~\ref{tab:simul_peak_fits_labconfig}. Along the non-focusing direction, the data are fitted again by a box profile, with center on $\mathrm{0.96 \pm 0.08}$~arcsec and with a width of $\mathrm{260 \pm 7}$~arcsec.
Interestingly, along the Z direction, which is less affected by assembly errors, the width of the simulated realistic image is perfectly consistent with the width of the real image of the prototype along the vertical direction.

\begin{table}
    \centering
        \caption{Fit parameters of the six peaks forming the combined image of the 12  Ge(220) crystals composing the simulated sector, projected along the focusing direction, in the laboratory configuration.}
    \begin{tabular}{cccc}
    \\
    \toprule
         Peak Number & Normalization & Mean [arcsec] & FWHM [arcsec]\\
    \midrule 
         1 & $\mathrm{0.98 \pm 0.01}$ & $\mathrm{-242.5 \pm 0.2}$ & $\mathrm{49.7 \pm 0.7}$ \\

        2 & $\mathrm{0.284 \pm 0.005}$ & $\mathrm{-167.6 \pm 0.2}$ & $\mathrm{35.5 \pm 0.7}$ \\

        3 & $\mathrm{0.282 \pm 0.005}$ & $\mathrm{-54.0 \pm 0.2}$ & $\mathrm{36.7 \pm 0.7}$ \\

        4 & $\mathrm{0.344 \pm 0.005}$ & $\mathrm{130.7 \pm 0.2}$ & $\mathrm{35.7 \pm 0.5}$ \\

        5 & $\mathrm{0.88 \pm 0.01}$ & $\mathrm{218.5 \pm 0.2}$ & $\mathrm{37.5 \pm 0.6}$ \\

        6 & $\mathrm{0.259 \pm 0.004}$ & $\mathrm{273.3 \pm 0.2}$ & $\mathrm{37.0 \pm 0.7}$ \\
    \bottomrule 
    \end{tabular}
    \label{tab:simul_peak_fits_labconfig}
\end{table}

\subsection{Module of Laue lens: astrophysical configuration}
\label{sector}
To assess the impact of the beam divergence, we simulated the behaviour of the prototype in an astrophysical configuration, i.e. with a lens-detector distance of 20 m and with a point-like source at an infinite distance. Even in this case, we simulated both an ideal lens sector and a sector with the same curvature radius distribution of the crystals and assembly error distribution as the one we measured on the prototype.

\begin{figure}
    \centering
    \includegraphics[scale = 0.3]{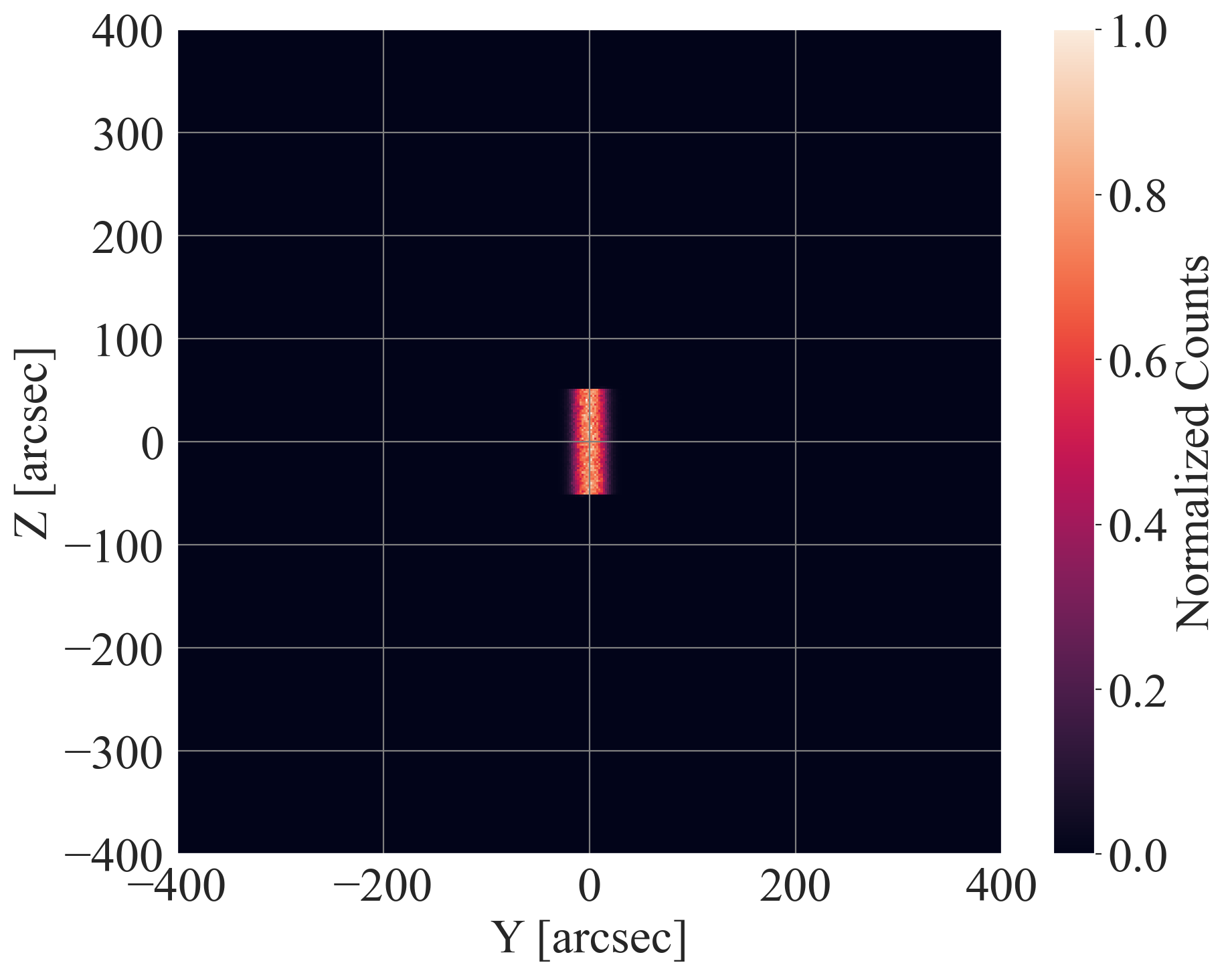}
    \includegraphics[scale = 0.3]{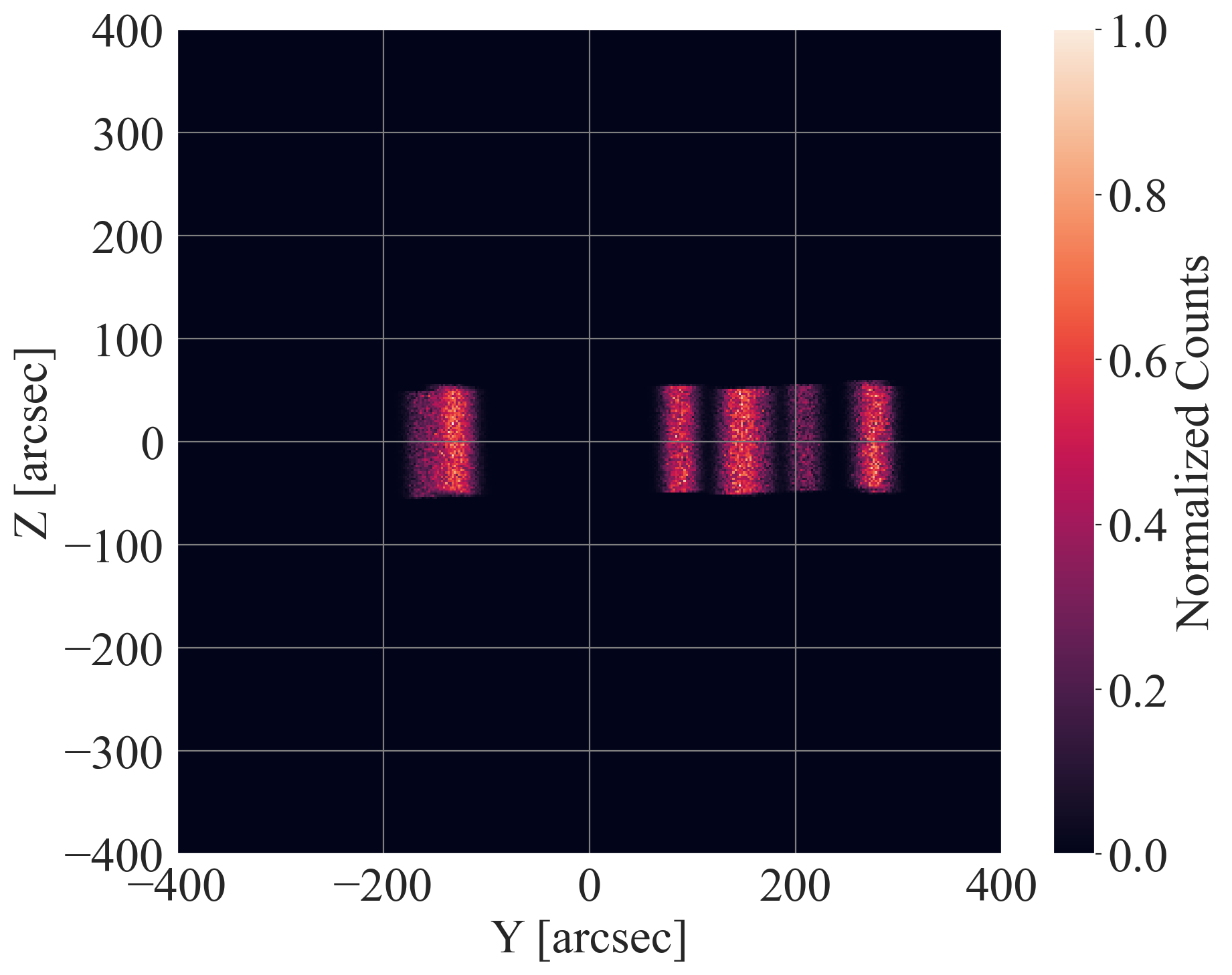}
    \includegraphics[scale = 0.3]{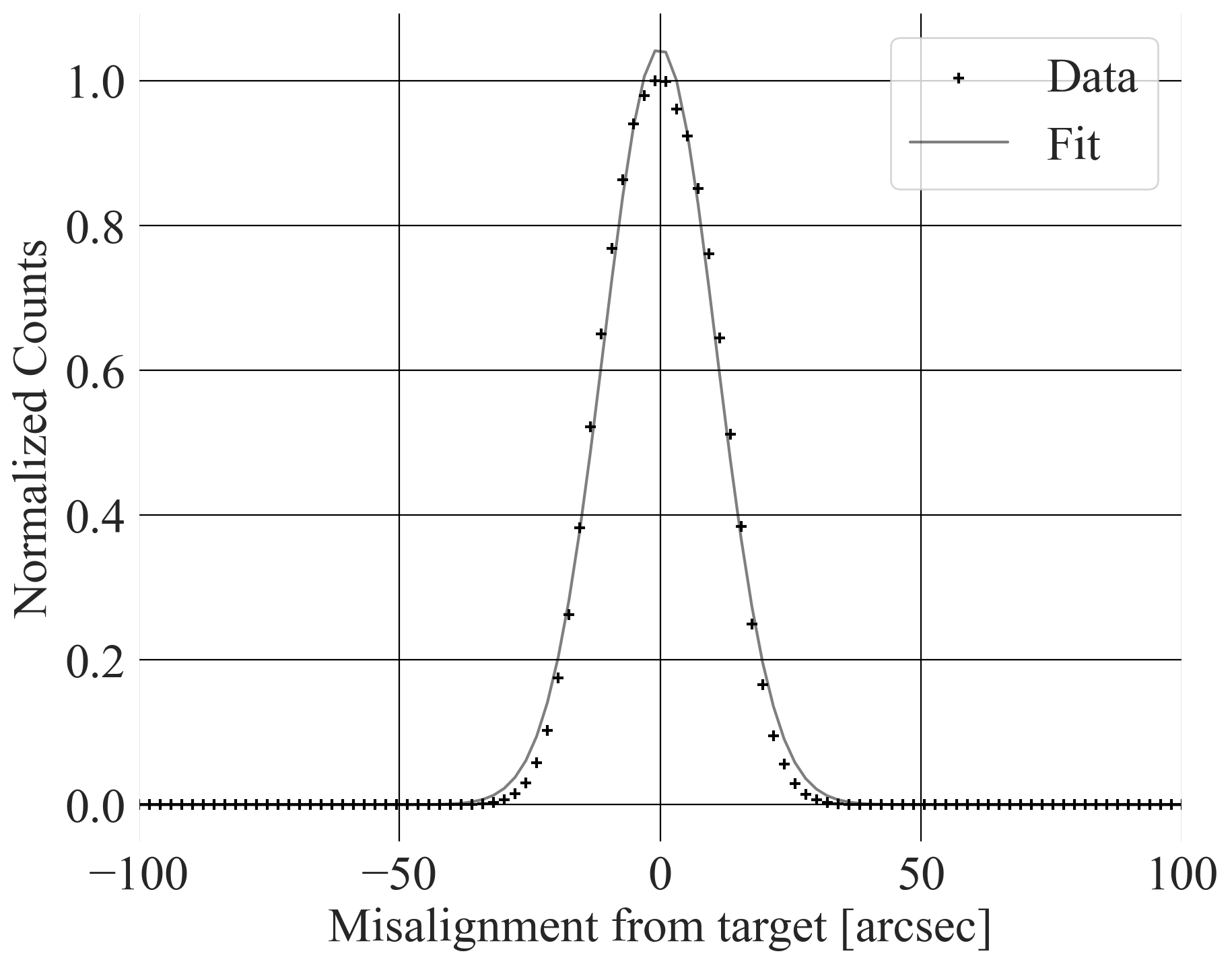}
    \includegraphics[scale = 0.3]{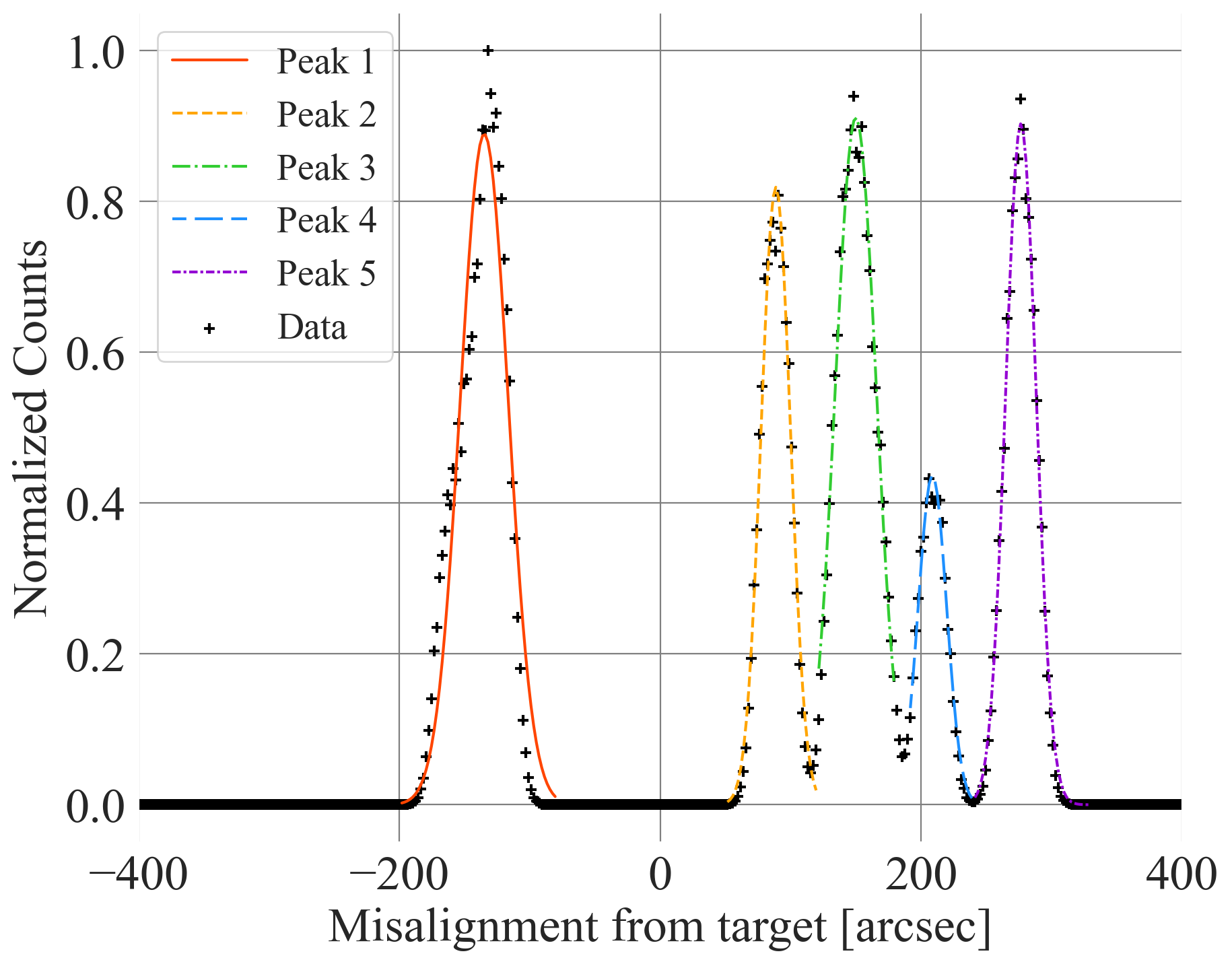}
    \includegraphics[scale = 0.3]{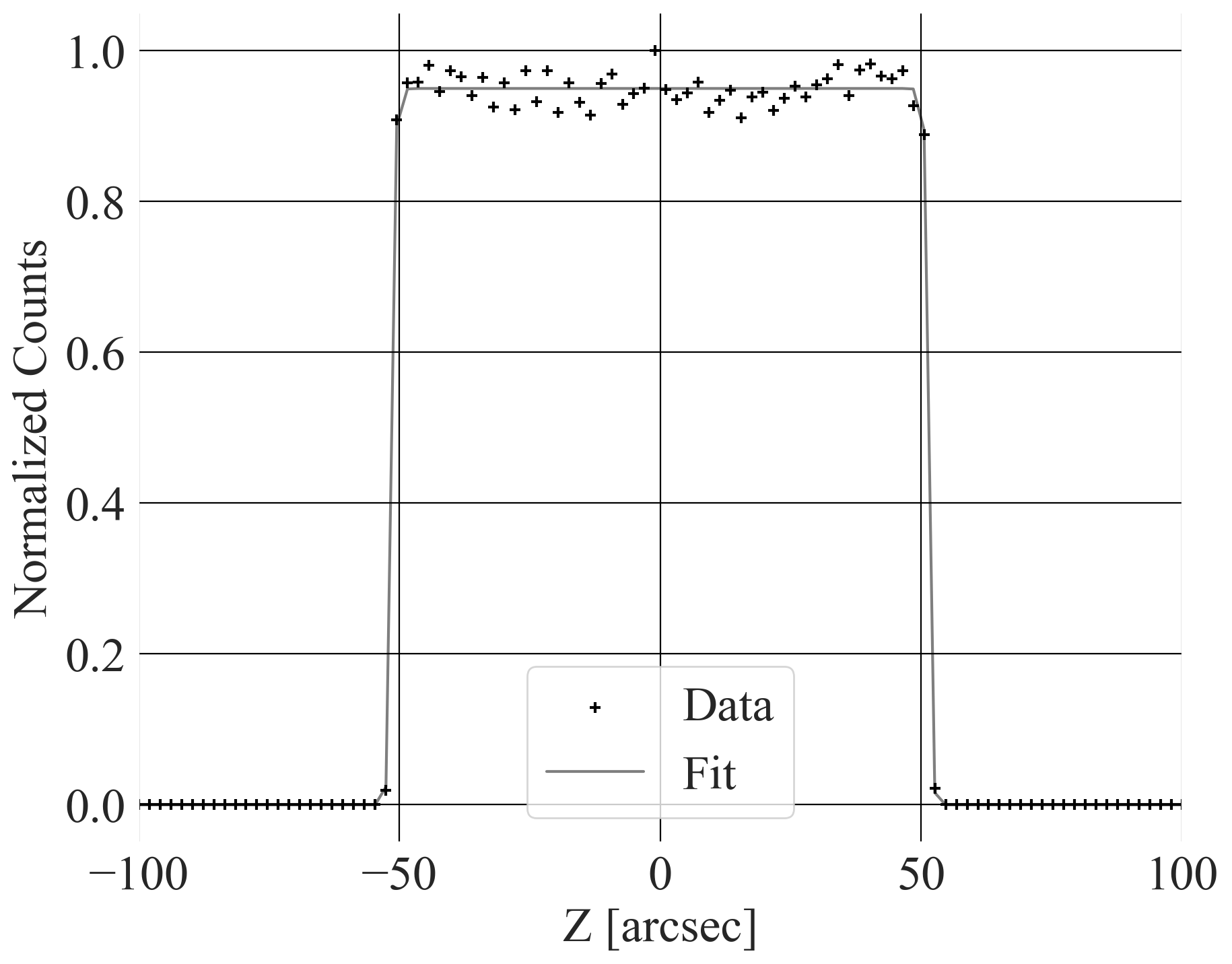}
    \includegraphics[scale = 0.3]{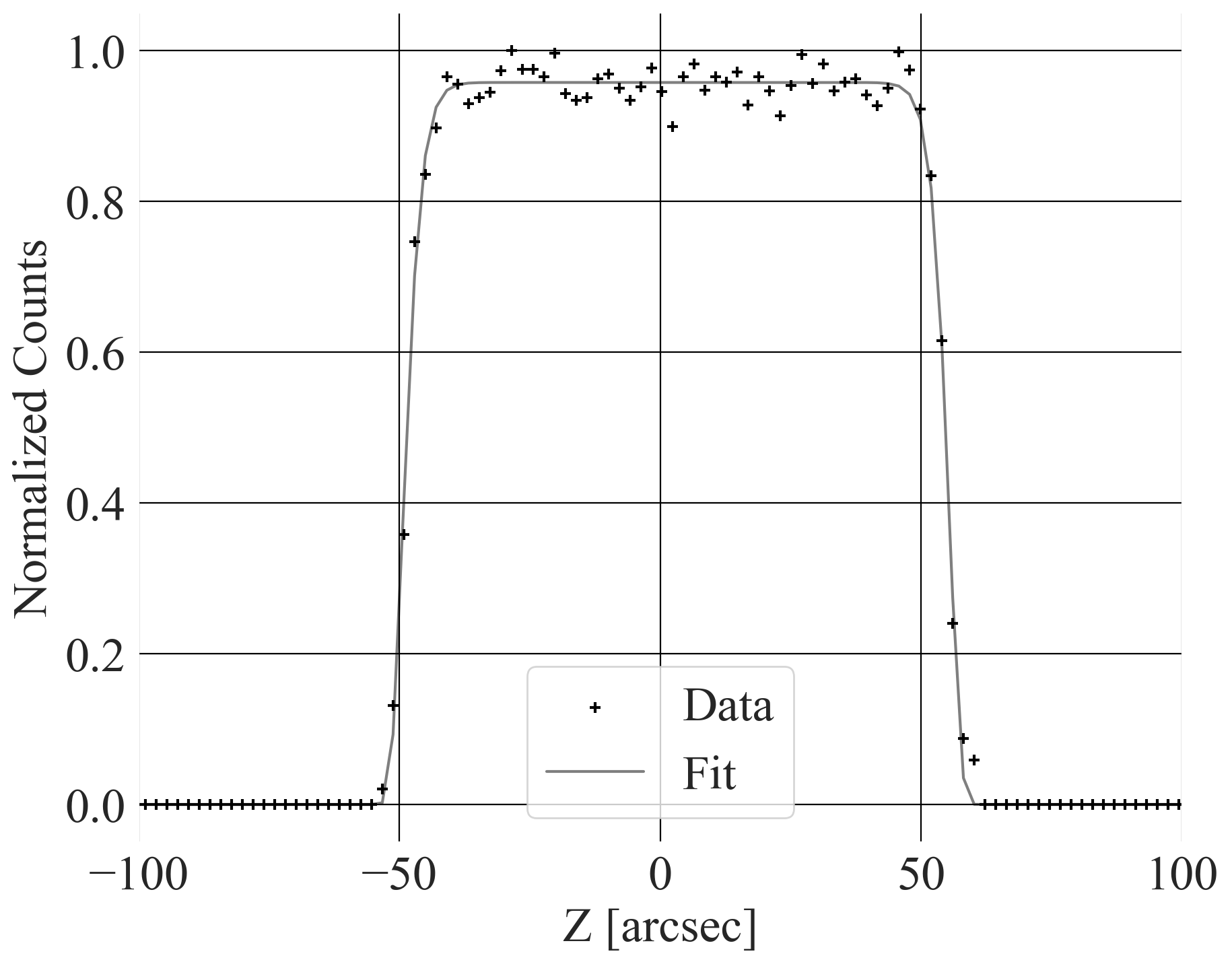}
    \caption{Comparison of the simulated images and profiles in the case of ideal sector configuration (left) and real configuration (right), with an astrophysical source. Top: Focal plane images of an ideal and a real sector. Center: Profiles along the focusing direction. Bottom: Profiles along the non-focusing direction.}
    \label{fig:simulations_sector_astro}
\end{figure}

In the ideal lens case, the profile of the combined image from the 12 crystals along the focusing direction has a Gaussian profile with a FWHM of 36.0~$\mathrm{\pm}$~0.1~arcsec.
Along the non-focusing direction, the data are fitted with a box profile centered on $\mathrm{-0.01 \pm 0.03}$~arcsec with width of $\mathrm{103.6 \pm 4}$~arcsec.
Images and profiles along Z and Y direction from both the simulated configurations are shown in Fig.~\ref{fig:simulations_sector_astro}.

In the realistic configuration, as a result of the random sampling of the misalignment distributions, the overall image of the crystals along the focusing direction shows five separate peaks. These peaks can be fitted with Gaussian profiles whose parameters are reported in Tab.~\ref{tab:simul_peak_fits}. Along the non-focusing direction, the data are fitted again by a box profile, with center on $\mathrm{3.05 \pm 0.03}$~arcsec and with a width of $\mathrm{103 \pm 4}$~arcsec. Again, this is compatible with the expected angular size of the 10 mm long non-focusing side, at a distance of 20 m, for a parallel beam.

\begin{table}
    \centering
        \caption{Fit parameters of the five peaks forming the combined image of the 12 Ge(220) crystals composing the simulated sector, projected along the focusing direction, in an astrophysical configuration (point-like source at infinity).}
    \begin{tabular}{cccc}
    \\
    \toprule
         Peak Number & Normalization & Mean [arcsec] & FWHM [arcsec]\\
    \midrule 
         1 & $\mathrm{0.89 \pm 0.03}$ & $\mathrm{-135.0 \pm 0.6}$ & $\mathrm{61 \pm 2}$ \\

        2 & $\mathrm{0.82 \pm 0.01}$ & $\mathrm{88.5 \pm 0.2}$ & $\mathrm{37.6 \pm 0.6}$ \\

        3 & $\mathrm{0.91 \pm 0.01}$ & $\mathrm{149.8 \pm 0.2}$ & $\mathrm{52.4 \pm 0.9}$ \\

        4 & $\mathrm{0.437 \pm 0.007}$ & $\mathrm{208.6 \pm 0.2}$ & $\mathrm{36.0 \pm 0.7}$ \\

        5 & $\mathrm{0.903 \pm 0.008}$ & $\mathrm{276.4 \pm 0.1}$ & $\mathrm{38.9 \pm 0.4}$ \\
    \bottomrule 
    \end{tabular}
    \label{tab:simul_peak_fits}
\end{table}

\subsection{Full Laue lens}
\label{full_lens}

\begin{figure}
    \centering
    \includegraphics[scale = 0.3]{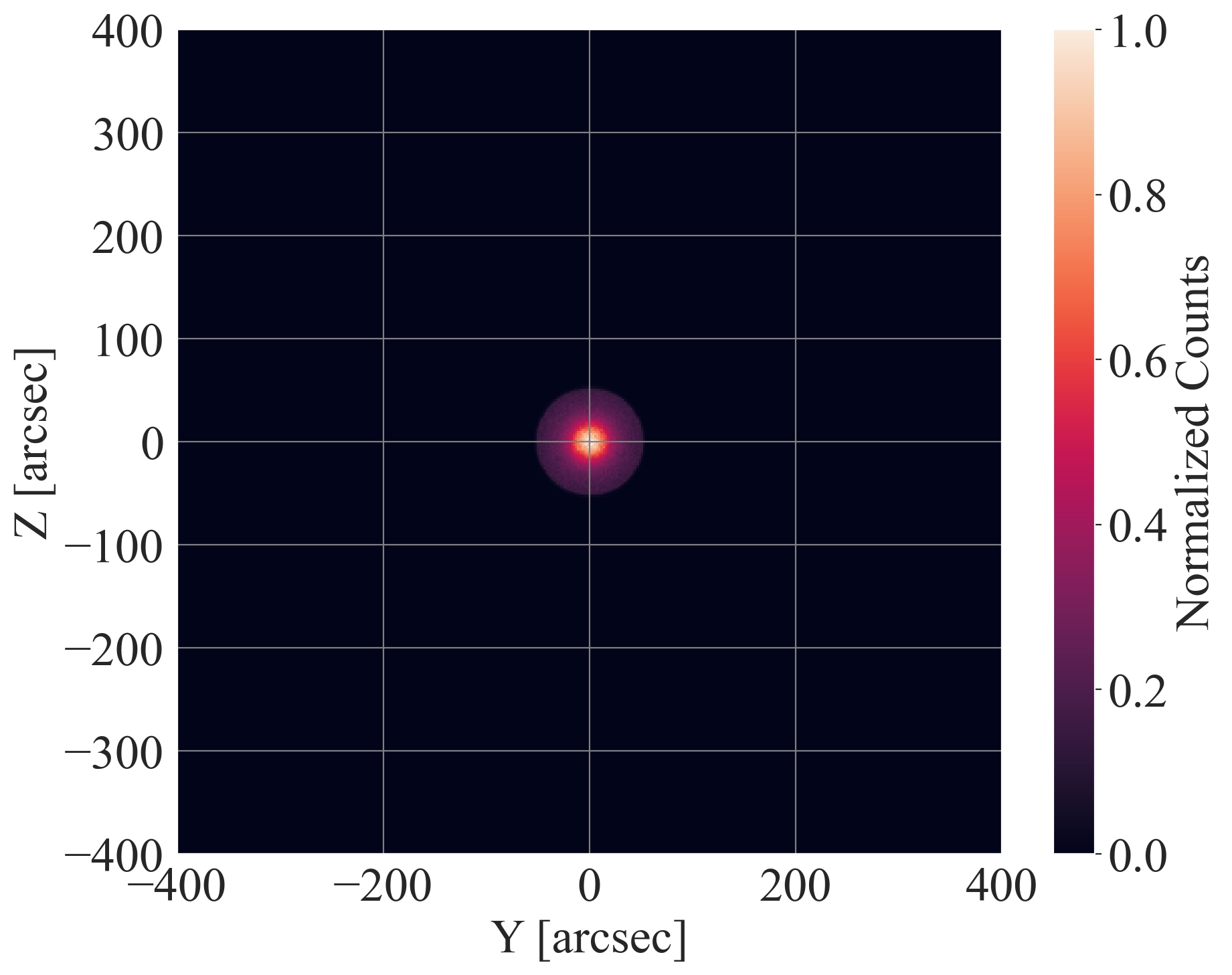}
    \includegraphics[scale = 0.3]{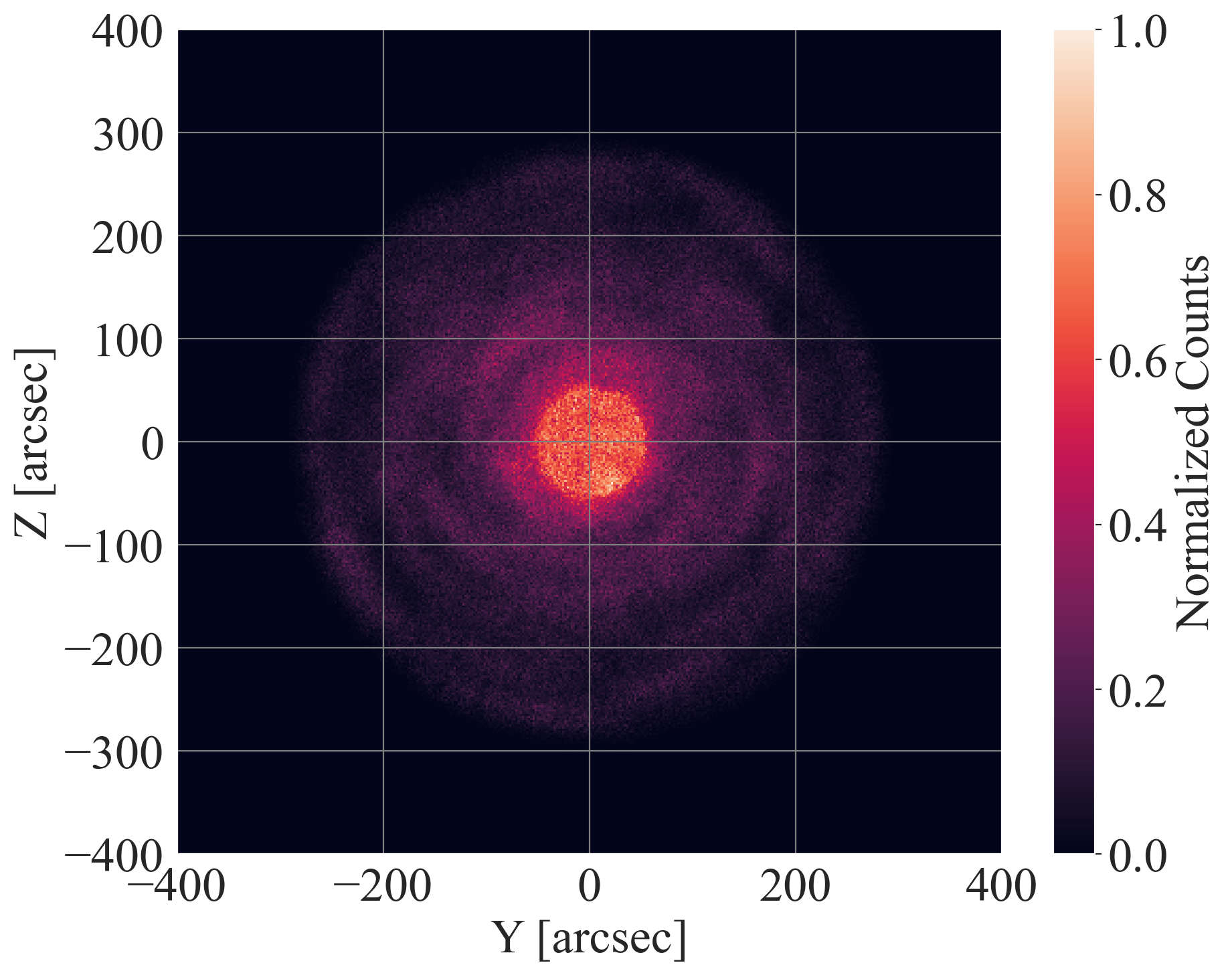}
    \includegraphics[scale = 0.27]{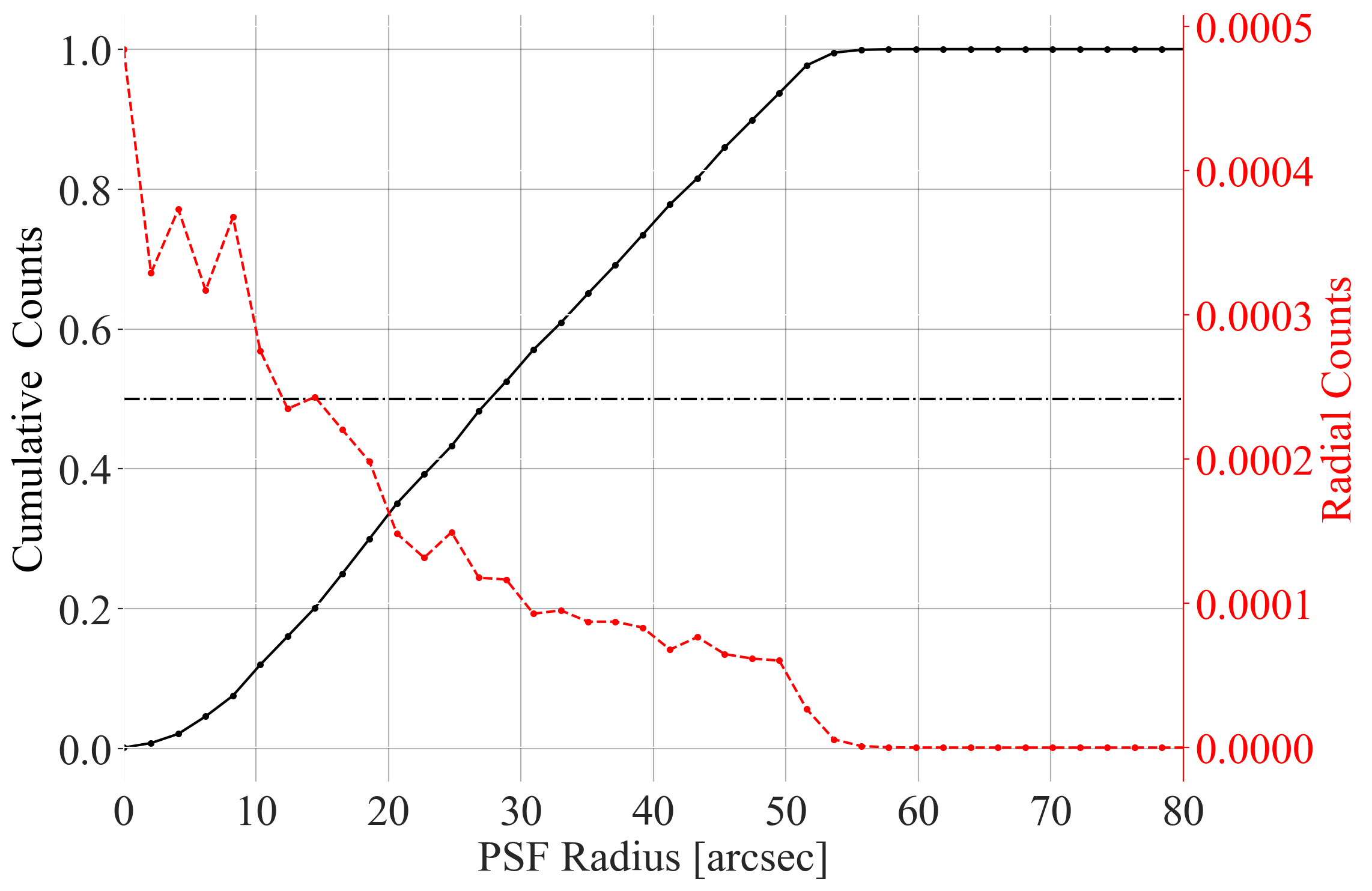}
    \includegraphics[scale = 0.27]{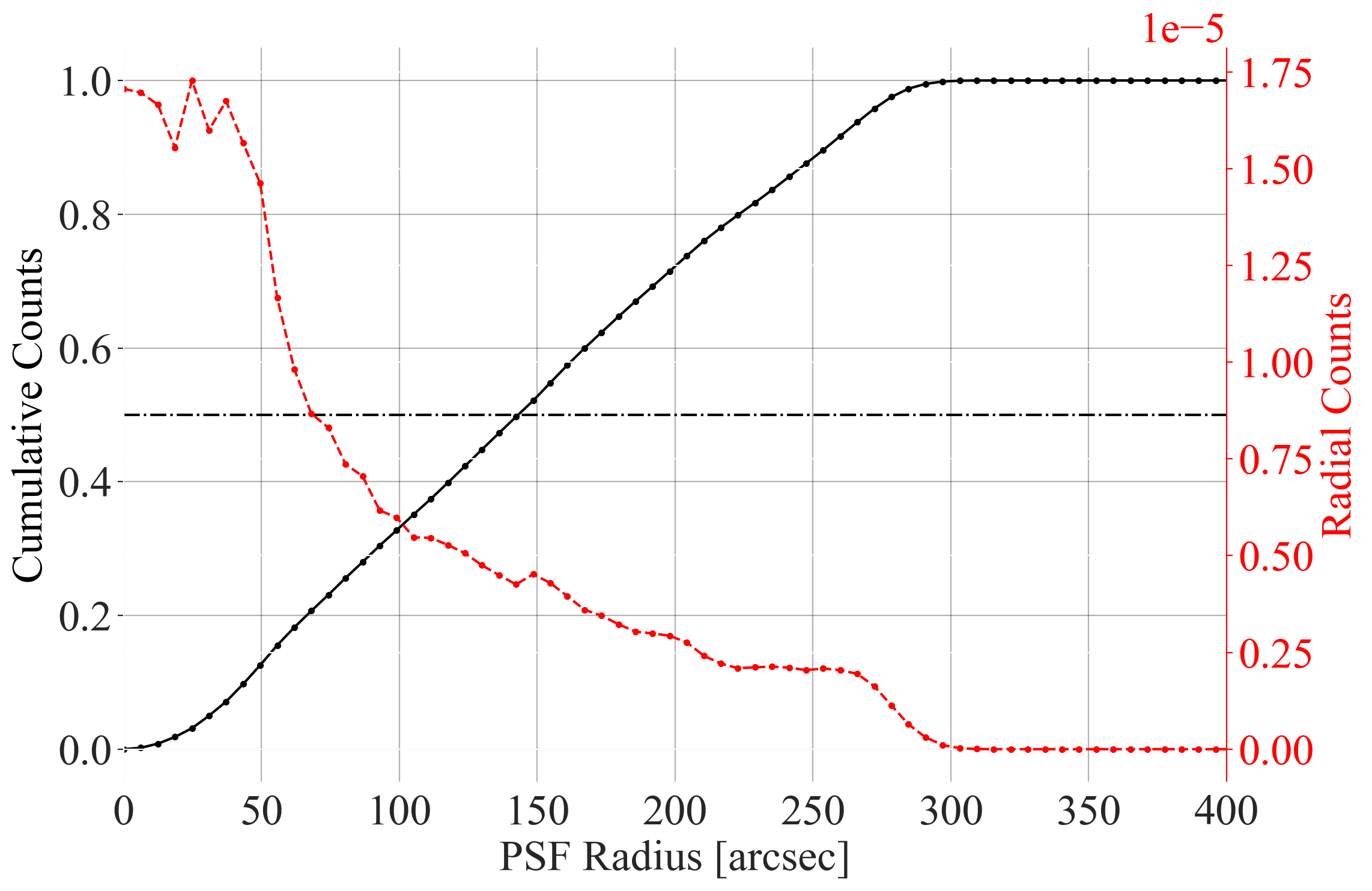}
    \caption{Left: (top) image produced by a simulated ideal full Laue lens built with Ge(220) crystals working in the 100-500 keV energy band; (bottom) radial (red curve) and cumulative (black curve) counts from the center of the PSF. All the crystals are bent with the same curvature radius of 40 m and no misalignment errors are present. The black dot-dashed line in the middle represents the 50\% integrated counts level. Right: same as in the left panel, but with a real Laue lens in which the curvature radius of the crystals are distributed according to the measured uniform distribution (center = $39.7$~m, width = $1.0$~m) and the misalignment errors are uniformly distributed as observed (center = 101 arcsec, width = 161 arcsec for the Bragg angles; center = 2 arcsec, width = 9 arcsec for the radial positioning angles). The black dot-dashed horizontal line represents the 50\% cumulative counts.
 }
    \label{fig:full_lens_sim_on_axis}
\end{figure}

With the same set of parameters characterizing the astrophysical configuration, we simulated a full Laue lens made of $\mathrm{\sim}$13700 Ge(220) crystals working in the energy range 100 - 500 keV, in the ideal case of no errors on the curvature radius of the crystals and no misalignment errors. Results from the simulations are shown in Fig.~\ref{fig:full_lens_sim_on_axis}, left. For the ideal configuration case, the Half Power Diameter (HPD) of the PSF is 54$\mathrm{\pm}$4~arcsec{, mainly due to the mosaicity of the crystals}. By adding a uniform distribution of alignment and curvature radius errors, reproducing the parameter spread measured on our prototype, we obtain the results in Fig.~\ref{fig:full_lens_sim_on_axis}, right. In this case, the PSF is broadened to a Half-Power diameter of 289$\mathrm{\pm}$4~arcsec.

\section{Discussion and conclusion}
\label{conclusion}
We built a prototype of a Laue lens sector made of 11 bent Ge(220) crystals bonded on a glass substrate by means of an UV-curable adhesive. From extensive testing, we obtained an average assembly misalignment in the Bragg's angle value of $50 \pm 25$~arcsec and an average misalignment in the polar angle position of $1.0\pm 1.5$~arcsec. We were able to reach a good accuracy in the radial positioning alignment, whereas the misalignment in the Bragg's angle is two orders of magnitudes larger. We observed that the position of the crystals is subjected to fluctuations in the first 3–4 days after the gluing procedure, however it settles afterward. 

Simulations based on our physical model of the Laue lens corroborate these observations and well reproduce the observed PSF of the diffracted image by a single sector, when the experimental setup is adopted, namely the geometrical configuration of the source-lens, the number of crystals and the alignment error distribution. 

By extrapolating these performances to a full Laue lens assembly, adopting a uniform distribution of misalignment errors, simulations show a final PSF with a HPD $\mathrm{= 289 \pm 4}$~arcsec.
This is the best result we have been able to achieve in building a long focal Laue lens prototype with adhesive-based bonding techniques. The Narrow Field Telescope on board the ASTENA mission has a PSF requirement of 30 arcsec, which would call for a reduction of at least one order of magnitude in the misalignment in the Bragg's angle  with respect to what we have obtained to date. The most significant source of misalignment in our process is the unpredictability of the volumetric shrinkage of the glue, which, combined with the release of heat during the cure, makes it difficult to position the crystal with the desired level of accuracy.

To address the challenges of the bonding technique outlined above, we are currently investigating various alternative approaches. The final goal is to achieve a robust bonding of the crystals while ensuring their precise  alignment. As previously emphasised, the role of the adhesive is critical, particularly when its thickness lacks uniformity, leading to uneven shrinkage during the curing phase. By reducing the miscut angle to a few arcseconds and, more importantly, by using crystals with a uniform miscut, 
one could achieve a significant reduction of the thickness of the adhesive between the crystals and the substrate, thus alleviating the stress induced by the curing process.
\newline
We are currently investigating alternatives that do not rely on traditional adhesives, specifically we are exploring either anodic or silicate bonding techniques,  which are widely used in the field of electronics.
\newline
We are also studying the potential application of small mechanisms based on piezoelectric actuators. This technique would have the advantage to enable both the alignment of the crystals (or modules of crystals) with arcsecond accuracy and the periodic realignment of the system, however the effect of the absorption of the material needed for the system must be evaluated. The latter is a feature not achievable with all bonding methods investigated thus far.
\newline
The research conducted so far shows that the level of positional accuracy of bent crystals achieved with UV-curable adhesives allows a long-focal length Laue lens to focus hard X-ray radiation onto a spot with a half-power diameter of 4.8 arcmin. With such a performance for the entire lens, even though it is still far from the 0.5-1 arcmin goal, one could build a narrow-field telescope for high energy spectro-polarimetry capable of obtaining a significant jump in sensitivity in the 50-600 keV band, when compared to current non-focusing instrumentation, thereby opening a new window in hard X and soft gamma-ray astrophysics.

\subsection*{Disclosures}
Authors have no conflict of interest to declare.

\subsection*{Code, Data, and Materials Availability}
The data and the simulation codes are accessible upon request to the authors.

\subsection*{Acknowledgments}
This work is partly supported by the AHEAD-2020 Project grant agreement 871158
of the European Union’s Horizon 2020 Program and by the ASI-INAF agreement
no. 2017-14-H.O ”Studies for future scientific missions”.

\bibliography{sn-bibliography}   

\begin{thebibliography}{10}

\bibitem{frontera2011}
F.~Frontera and P.~V. Ballmoos, ``Laue gamma-ray lenses for space astrophysics:status and prospects,''  (2011).

\bibitem{Virgilli2016}
E.~{Virgilli}, F.~{Frontera}, P.~{Rosati}, {\em et~al.}, ``{Focusing effect of bent GaAs crystals for {\ensuremath{\gamma}}-ray Laue lenses: Monte Carlo and experimental results},'' {\em Experimental Astronomy} {\bf 41}, 307--326  (2016).

\bibitem{Frontera2021}
F.~Frontera, E.~Virgilli, C.~Guidorzi, {\em et~al.}, ``Understanding the origin of the positron annihilation line and the physics of supernova explosions,'' {\em Experimental Astronomy} {\bf 51}, 1175--1202  (2021).

\bibitem{Guidorzi2021}
C.~Guidorzi, F.~Frontera, G.~Ghirlanda, {\em et~al.}, ``A deep study of the high--energy transient sky,'' {\em Experimental Astronomy} {\bf 51}, 1203--1223  (2021).

\bibitem{Ferrari2013}
C.~{Ferrari}, E.~{Buffagni}, E.~{Bonnini}, {\em et~al.}, ``{X-ray diffraction efficiency of bent GaAs mosaic crystals for the LAUE project},'' in {\em Optics for EUV, X-Ray, and Gamma-Ray Astronomy VI},  S.~L. {O'Dell} and G.~{Pareschi}, Eds.,  {\bf 8861}, 568 -- 575, SPIE  (2013).

\bibitem{Buffagni2015}
E.~{Buffagni}, E.~{Bonnini}, C.~{Ferrari}, {\em et~al.}, ``{X-ray characterization of curved crystals for hard x-ray astronomy},'' in {\em EUV and X-ray Optics: Synergy between Laboratory and Space IV},  R.~{Hudec} and L.~{Pina}, Eds., {\em Society of Photo-Optical Instrumentation Engineers (SPIE) Conference Series} {\bf 9510}  (2015).

\bibitem{Authier98}
A.~Authier and C.~Malgrange, ``{Diffraction Physics},'' {\em Acta Crystallographica Section A} {\bf 54}, 806--819  (1998).

\bibitem{Virgilli2014}
E.~{Virgilli}, F.~{Frontera}, V.~{Valsan}, {\em et~al.}, ``{The LAUE project and its main results},'' {\em arXiv e-prints} , arXiv:1401.4948  (2014).

\bibitem{Ferro2022}
L.~{Ferro}, E.~{Virgilli}, M.~{Moita}, {\em et~al.}, ``{The TRILL project: increasing the technological readiness of Laue lenses},'' in {\em Space Telescopes and Instrumentation 2022: Ultraviolet to Gamma Ray},  J.-W.~A. {den Herder}, S.~{Nikzad}, and K.~{Nakazawa}, Eds., {\em Society of Photo-Optical Instrumentation Engineers (SPIE) Conference Series} {\bf 12181}, 121812K  (2022).

\bibitem{Virgilli17}
E.~Virgilli, V.~Valsan, F.~Frontera, {\em et~al.}, ``{Expected performances of a Laue lens made with bent crystals},'' {\em Journal of Astronomical Telescopes, Instruments, and Systems} {\bf 3}(4), 044001  (2017).

\end{thebibliography}
\bibliographystyle{spiejour}

\vspace{2ex}\noindent\textbf{Lisa Ferro} is a Ph.D. student in Physics at the University of Ferrara (UniFe), whose work is supervised by P. Rosati, F. Frontera, C. Guidorzi from UniFe, and E. Virgilli from INAF-OAS of Bologna. Her main research topics are the development of focusing optics for hard X and Gamma-rays, especially Laue lenses, and instrumentation for high energy astrophysics.  

\end{spacing}
\end{document}